\documentclass[reprint,superscriptaddress,nofootinbib,amsmath,amssymb,aps,prl,floatfix]{revtex4-2}
\usepackage{booktabs}
\usepackage{graphicx}
\usepackage{dcolumn}
\usepackage{bm}
\usepackage{orcidlink}
\usepackage{multirow}
\usepackage{mathtools}
\renewcommand{\arraystretch}{1.2}
\newcommand{\ra}[1]{\renewcommand{\arraystretch}{#1}}

\begin{document}

\title{Quantum Enhanced Balanced Heterodyne Readout for Differential Interferometry}

\author{Daniel W. Gould \orcidlink{0000-0002-2915-4690}}
\email[Corresponding author: ]{daniel.gould@anu.edu.au}
\affiliation{OzGrav, Centre for Gravitational Astrophysics, Research School of Physics and Research School of Astronomy and Astrophysics, Australian National University, Australian Capital Territory, Australia}
\author{Vaishali B. Adya \orcidlink{0000-0003-4955-6280}}
\affiliation{Nonlinear and Quantum Photonics Lab, Department of Applied Physics, KTH Royal Institute of Technology, Stockholm, Sweden} 
\author{Sheon S. Y. Chua \orcidlink{0000-0001-8026-7597}}
\affiliation{OzGrav, Centre for Gravitational Astrophysics, Research School of Physics and Research School of Astronomy and Astrophysics, Australian National University, Australian Capital Territory, Australia}
\author{Jonas Junker \orcidlink{0000-0002-3051-4374}}
\affiliation{OzGrav, Centre for Gravitational Astrophysics, Research School of Physics and Research School of Astronomy and Astrophysics, Australian National University, Australian Capital Territory, Australia}
\affiliation{Institute for Gravitational Physics, Leibniz University Hannover, and Max Planck Institute for Gravitational Physics (Albert Einstein Institute), Hannover, Germany}
\author{Dennis Wilken \orcidlink{0000-0002-7290-9411}}
\affiliation{Institute for Gravitational Physics, Leibniz University Hannover, and Max Planck Institute for Gravitational Physics (Albert Einstein Institute), Hannover, Germany}
\author{Terry G. McRae \orcidlink{0000-0002-6540-6824}}
\affiliation{OzGrav, Centre for Gravitational Astrophysics, Research School of Physics and Research School of Astronomy and Astrophysics, Australian National University, Australian Capital Territory, Australia}
\author{Bram J. J. Slagmolen \orcidlink{0000-0002-2471-3828}}
\affiliation{OzGrav, Centre for Gravitational Astrophysics, Research School of Physics and Research School of Astronomy and Astrophysics, Australian National University, Australian Capital Territory, Australia}
\author{Min Jet Yap \orcidlink{0000-0002-6492-9156}}
\affiliation{OzGrav, Centre for Gravitational Astrophysics, Research School of Physics and Research School of Astronomy and Astrophysics, Australian National University, Australian Capital Territory, Australia}
\author{Robert L. Ward \orcidlink{0000-0001-5503-5241}}
\affiliation{OzGrav, Centre for Gravitational Astrophysics, Research School of Physics and Research School of Astronomy and Astrophysics, Australian National University, Australian Capital Territory, Australia}
\author{Mich\`{e}le Heurs \orcidlink{0000-0002-5577-2273}}
\affiliation{Institute for Gravitational Physics, Leibniz University Hannover, and Max Planck Institute for Gravitational Physics (Albert Einstein Institute), Hannover, Germany}
\author{David E. McClelland \orcidlink{0000-0001-6210-5842}}
\affiliation{OzGrav, Centre for Gravitational Astrophysics, Research School of Physics and Research School of Astronomy and Astrophysics, Australian National University, Australian Capital Territory, Australia}

\begin{abstract}
  Conventional heterodyne readout schemes are now under reconsideration due to the realization of techniques to evade its inherent $3\;{\rm dB}$ signal-to-noise penalty. The application of high-frequency, quadrature-entangled, two-mode squeezed states can further improve the readout sensitivity of audio-band signals. In this Letter, we experimentally demonstrate quantum-enhanced heterodyne readout of two spatially distinct interferometers with direct optical signal combination, circumventing the $3\;{\rm dB}$ heterodyne signal-to-noise penalty. Applying a high-frequency, quadrature-entangled, two-mode squeezed state, we show further signal-to-noise improvement of an injected audio band signal of $3.5\;{\rm dB}$. This technique is applicable for quantum-limited high-precision experiments, with application to searches for quantum gravity, searches for dark matter, gravitational wave detection, and wavelength-multiplexed quantum communication.
\end{abstract}
\maketitle
Quantum enhancement techniques, such as the application of squeezed states of light, have improved the quantum-limited signal-to-noise ratio (SNR) in many precision measurements, including gravitational wave detection~\cite{Ligo_QResp,Ganapathy2023,PhysRevLett.123.231108,PhysRevLett.126.041102}, magnetometry~\cite{PhysRevLett.127.193601}, biosensing~\cite{Bowen21}, and plasmonics~\cite{PhysRevA.92.053812}. Squeezed states are also a major interest for quantum communication~\cite{armstrong2012programmable}. The long-developed ubiquitous methods of homodyne detection~\cite{Collett} and heterodyne detection~\cite{Yuen:83} are critical to the success of applying squeezed states. Balanced homodyne detection, where the signal sidebands are symmetric around the local oscillator (LO) field, continues to be applied in many quantum metrology areas, including measuring highly squeezed states~\cite{PhysRevLett.129.121103} and future gravitational wave detectors~\cite{Fritschel:14}. However, for low-frequency sensitivity, care is needed to avoid technical noise sidebands on the LO field affecting the measurement~\cite{Stefszky_2012,Steinlechner2015}.

To circumvent the technical challenges of producing, controlling, and utilizing low (or audio-band) frequency squeezed light~\cite{Stefszky_2012,McKenzie2004, Vahlbruch2007,PhysRevLett.97.011101}, several theoretical techniques have been proposed revolving around heterodyne readout schemes~\cite{Yuen:83}. Heterodyning is commonplace in several experiments such as light detection and ranging (LIDAR)~\cite{Spollard:21}, vibrometry~\cite{Yu:23}, spectroscopy~\cite{deAndrade:20}, and laser stabilization~\cite{PDH}. Heterodyning offers an essential advantage of innate resistance to low-frequency technical noise on the LO due to the significant frequency separation between it and the signals of interest. The sideband structure of a typical heterodyne experiment has a single-sided signal sideband and dual-sided vacuum noise sidebands~\cite{Yuen:83}. Compared to homodyne detection techniques, where both signal and noise have a dual-sided sideband structure, heterodyne detection suffers from a $3\;{\rm dB}$ reduction in the overall SNR.

This SNR penalty can be evaded with a two-carrier approach, introducing coherent signals at positive and negative frequencies relative to the LO~\cite{Collett,PRL_1D_OL}. Further SNR improvement is possible using two-mode squeezed states~\cite{Xie:18}. Avoiding the LO noise coupling and challenges of low-frequency squeezing makes two-carrier heterodyning an attractive alternative to homodyne detection for quantum-enhanced interferometry-based sensors~\cite{Zhang_two_carrier,JZ_two_carrier}. Additionally, two-carrier heterodyning can provide an all-optical differential readout method for twin sensors, such as those used for tests of quantum gravity in correlated space-time volumes~\cite{Vermeulen_2021,TheHolometer,Pradyumna2020} and searches for dark matter candidates~\cite{DMPolarimetryEjlli,LIDAFirstResults}.

Applying a two-mode squeezed state to a twin interferometer can increase the SNR, which has been theoretically investigated for gravitational wave detectors~\cite{Zhang_two_carrier,JZ_two_carrier}, and interferometric tests of quantum gravity~\cite{QL_coupled_int,one_two_SQ_corr_int}. Two-mode squeezed states have been experimentally investigated in application to twin-interferometer~\cite{Pradyumna2020} and single interferometer designs~\cite{EPR_Sudbeck,EPR_Yap,Gould_QNC}. These works rely on the use of homodyne readout and digital filtering.

In this Letter, we experimentally demonstrate a quantum-enhanced heterodyne readout of two spatially distinct interferometers with direct optical signal combination. We circumvent the 3 dB heterodyne SNR penalty, showing further SNR improvements by applying two-mode squeezing generated with a frequency-nondegenerate optical parametric oscillator (OPO). We provide an alternative to the conventional homodyne readout without compromising on sensitivity, an approach that can reduce the complexity of real-time quantum communication experiments~\cite{PhysRevA.97.032329,Qtele}. Our scheme is of particular interest for correlated interferometry experiments~\cite{TheHolometer,Holometer1stmeas, QL_coupled_int,one_two_SQ_corr_int,Pradyumna2020,Vermeulen_2021}, dark matter detectors~\cite{DMPolarimetryEjlli,LIDAFirstResults}, and gravitational wave detectors~\cite{Zhang_two_carrier,JZ_two_carrier}, allowing flexibility in the choice of readout quadratures and quadrature combinations. In application to these detectors, this readout scheme may provide access to signal- and null-data streams simultaneously. Using a frequency-nondegenerate OPO to produce two-mode squeezing minimizes technical complexity compared to the widely used method in quantum teleportation~\cite{Qtele_Science} (via optical combination of two degenerate squeezed states), as highlighted in~\cite{Chapman:23}. Additionally, our method of squeezed state generation reduces inherent losses compared to previous twin-interferometer demonstrations~\cite{Pradyumna2020}.

\begin{figure}[t]
  \centering
  \includegraphics[width=\columnwidth]{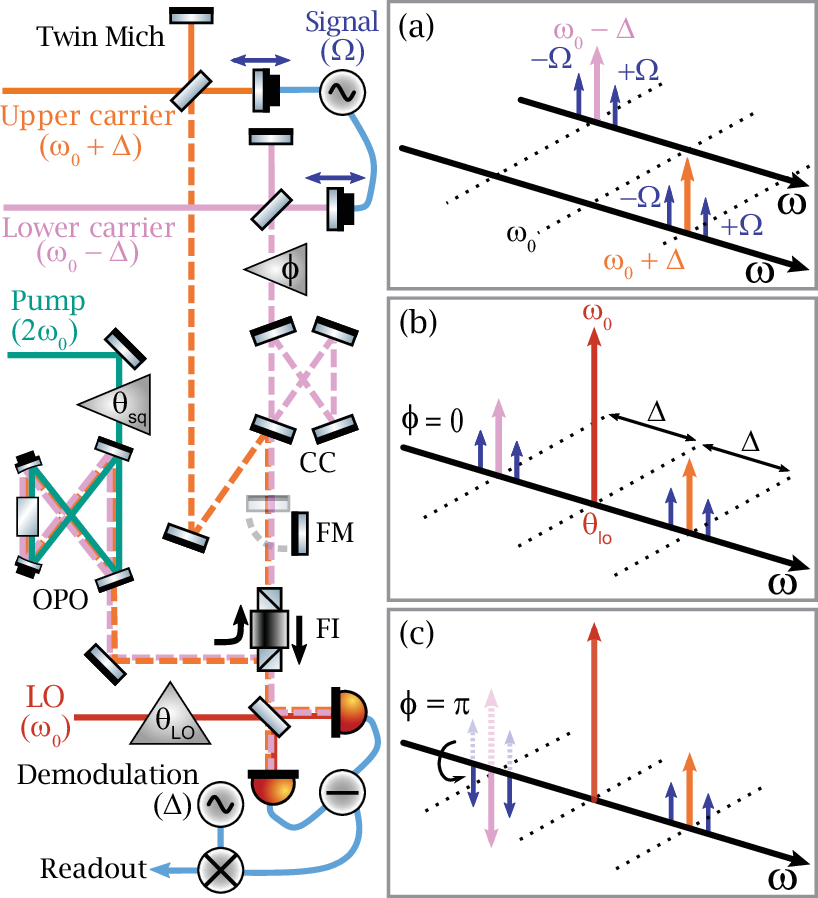}
  \caption{Simplified schematic of the experiment, showing the twin-Michelson interferometer (Twin Mich) outputs that are spatially overlapped with a combining cavity (CC), and phase locked (relative phase $\phi$). The combined fields pass through a Faraday isolator (FI) towards the balanced detector. A two-mode squeezed state, produced by the optical parametric oscillator (OPO), is injected via the Faraday isolator, with each mode being reflected off an interferometer. A flipper mirror (FM) is used for direct squeezing measurement. Michelson carrier field preparation and polarization auxiliary fields are not shown for clarity. The LO phase ($\theta_{\rm LO}$) and squeezed quadrature angle ($\theta_{\rm sq}$) are also shown.~(a) Frequency diagram of each interferometer carrier with $\pm\,\Omega$ signal sidebands.~(b) and (c) Frequency diagrams of fields at the balanced detector with in-phase ($\phi = 0$) and out-of-phase ($\phi=\pi$) carriers.}\label{Fig:1}
\end{figure}

The experimental setup is illustrated in Fig.~\ref{Fig:1}. Two spatially separate Michelson interferometers are probed with $40\;{\rm uW}$ \textit{s}-polarized upper ($\omega_{0}+\Delta$) and lower ($\omega_{0}-\Delta$) carrier fields, frequency separated by $2\Delta = 850\;{\rm MHz}$, and operated close to the dark fringe. We generate the two carrier fields by strongly modulating a portion of the primary $1064\;{\rm nm}$ laser field with an electro-optic modulator (EOM), driven at $\Delta = 425\;{\rm MHz}$, then spatially separated for use in the interferometers. The carrier field preparation is not shown in Fig.~\ref{Fig:1}. A signal ($\Omega = 25\;{\rm kHz}$) is introduced into each interferometer through piezoelectric-transducer driven end mirrors, resulting in a frequency diagram as shown in Fig.~\ref{Fig:1}(a). After spatial recombination on the combining cavity, the interferometers' output fields are simultaneously measured.

\textit{P}-polarized auxiliary control fields (not shown in Fig.~\ref{Fig:1}) are generated with birefringent optics before each interferometer. Further birefringent optics are used to break the degeneracy of \textit{s} and \textit{p} polarization in the interferometers. This allows us to lock the interferometers with a \textit{p}-polarized field on the mid-fringe in reflection while simultaneously ensuring a dark fringe for the \textit{s}-polarization in transmission. The combining cavity is length controlled using the lower carrier auxiliary field, sensed on reflection. The frequency nondegeneracy between \textit{s} and \textit{p} polarization of the cavity is negligible by design. The auxiliary \textit{p}-polarization fields are reflected from the Faraday isolator, and their beat note forms the control signal between the two carriers and fed back to a PZT-actuated mirror, thus setting $\phi$.

The carrier fields are measured using balanced detection, after combination with the LO (at $\omega_{0}$) on a 50 : 50 beam splitter. Frequency diagrams of the fields after the beam splitter are shown in Fig.~\ref{Fig:1}(b) and ~\ref{Fig:1}(c). The beam splitter outputs are then sent to a high-bandwidth balanced detector (similar to the one used in~\cite{Junker2022}). Our readout scheme electronically demodulates the subtracted photocurrent at $\Delta = 425\;{\rm MHz}$, extracting the audio-frequency signals that modulate the high-frequency LO-carrier beat notes. The LO field is stabilized to a $550\;{\rm kHz}$ amplitude modulation sideband on the upper carrier. This beat note at the balanced detector provides a control signal to a  PZT-actuated mirror on the LO path, setting the LO phase $\theta_{\rm LO}=0$. Control of both $\theta_{\rm LO}$ and $\phi$ gives the choice of coherent signal addition or cancellation. 

Electronically demodulating the heterodyne's subtracted photocurrent output yields information about the sum and difference of the two interferometers quadratures~\cite{Qtele}. Defining the amplitude quadrature, $\Tilde{X}_{1}$, and phase quadrature, $\Tilde{X}_{2}$, of the upper and lower interferometers as $\Tilde{X}_{(1,2)}^{\pm\Delta}=\Tilde{X}_{(1,2)}(\omega_0\pm\Delta)$, the in-phase photocurrent $\Tilde{i}_I$ is proportional to:
\begin{eqnarray}
  \Tilde{i}_I(\theta_{\rm LO}=0,\phi=0)&\propto&\frac{1}{2}\left(\Tilde{X}_{1}^{-\Delta}+\Tilde{X}_{1}^{+\Delta}\right),\label{eqn:sumamp}\\
  \Tilde{i}_I(\theta_{\rm LO}=0,\phi=\pi)&\propto&\frac{1}{2}\left(\Tilde{X}_{1}^{-\Delta}-\Tilde{X}_{1}^{+\Delta}\right).\label{eqn:diffamp}
\end{eqnarray}
Conversely, when demodulating in-quadrature, the resulting photocurrent, $\Tilde{i}_Q$, contains information about the difference or sum of the phase quadratures of the two interferometers:
\begin{eqnarray}
  \Tilde{i}_Q(\theta_{\rm LO}=0,\phi=0)&\propto&\frac{1}{2}\left(\Tilde{X}_2^{-\Delta}-\Tilde{X}_2^{+\Delta}\right),\label{eqn:diffpha}\\
  \Tilde{i}_Q(\theta_{\rm LO}=0,\phi=\pi)&\propto&\frac{1}{2}\left(\Tilde{X}_2^{-\Delta}+\Tilde{X}_2^{+\Delta}\right).\label{eqn:sumpha}
\end{eqnarray}
Changes to $\theta_{\rm LO}$ and $\phi$ result in a mixing of the carrier quadratures. The demodulated photocurrents in Eqs.~(\ref{eqn:sumamp})-(\ref{eqn:sumpha}) can be generalized with the substitutions $\Tilde{X}_{(1,2)}^{-\Delta}\to\cos(\theta_{\rm LO})\Tilde{X}_{(1,2)}^{-\Delta}+\sin(\theta_{\rm LO})\Tilde{X}_{(2,1)}^{-\Delta}$ and $\Tilde{X}_{(1,2)}^{+\Delta}\to\cos(\phi-\theta_{\rm LO})\Tilde{X}_{(1,2)}^{+\Delta}+\sin(\phi-\theta_{\rm LO})\Tilde{X}_{(2,1)}^{+\Delta}$.

For this work, we focus on analyzing the amplitude quadrature signal dynamics [Eqs.~(\ref{eqn:sumamp}) and (\ref{eqn:diffamp})], motivated by the preferential coupling of differential arm-length modulation to amplitude modulation in interferometers at a near-dark fringe. For signals preferentially appearing in one quadrature, setting incorrect $\theta_{\rm LO}$ or $\phi$ will result in loss of signal. For the remainder of this Letter, $\theta_{\rm LO}$ is always set to zero.

A quantum-enhanced SNR can be achieved via the injection of a two-mode squeezed state, produced by a doubly resonant bow-tie OPO~\cite{StefszkyJOPB2011}. The two modes of the squeezed state are quadrature-entangled~\cite{QNOPO_Reid} and frequency separated by the OPO's free spectral range (FSR), $850\;{\rm MHz}$. This is also the chosen frequency separation of the interferometer carrier fields ($2\Delta$). The squeezed state is injected towards the interferometers with the Faraday isolator. The combining cavity spatially separates the two-mode squeezed state, with one mode entering the antisymmetric port of each interferometer. Upon reflection off each interferometer, the entangled modes copropagate with the upper and lower signal fields. They are spatially recombined at the combining cavity, and then measured with the balanced detector.

The quadrature-entangled squeezed state produced by our OPO contains correlations between the two amplitude quadratures and two phase quadratures of the modes. This state can be described by the relationship between each mode's amplitude and phase quadratures, $\Tilde{S}_{1}^{\pm\Delta}$ and $\Tilde{S}_{2}^{\pm\Delta}$~\cite{ExpCriteriaCV_Bowen, QNOPO_Reid}. The noise variance of the combined quadratures of a two-mode amplitude-quadrature squeezed state are
\begin{align}
  \frac{1}{2}{\rm Var}(\Tilde{S}_{1}^{-\Delta}+\Tilde{S}_{1}^{+\Delta})=\frac{1}{2}{\rm Var}(\Tilde{S}_{2}^{-\Delta}-\Tilde{S}_{2}^{+\Delta})=V_{-},\label{eqn:comb-sq}\\
  \frac{1}{2}{\rm Var}(\Tilde{S}_{1}^{-\Delta}-\Tilde{S}_{1}^{+\Delta})=\frac{1}{2}{\rm Var}(\Tilde{S}_{2}^{-\Delta}+\Tilde{S}_{2}^{+\Delta})=V_{+}.\label{eqn:comb-asq}
\end{align}
Where $V_{-}$ is the squeezed, and $V_{+}$ is the antisqueezed, noise variance:
\begin{equation}
  V_{-}=1-\frac{4x}{{(1+x)}^2}=e^{-2r},\;V_{+}=1+\frac{4x}{{(1- x)}^2}=e^{2r}.
\end{equation}
The normalized down-conversion parameter $x$ is experimentally measured as the ratio of pump power to the OPO's oscillation threshold power, $x=\sqrt{P/P_T}$. This is related to the squeezing parameter $r$ by $x=-\tanh(\frac{r}{2})$.

Equations~(\ref{eqn:comb-sq}) and (\ref{eqn:comb-asq}) show that measurements of a two-mode amplitude-quadrature squeezed state aligned to the LO, $\theta_{\rm sq}=0$, result in a squeezed noise variance in both simultaneously accessible demodulated photocurrents, Eqs.~(\ref{eqn:sumamp}) and (\ref{eqn:diffpha})~\cite{Qtele,ExpCriteriaCV_Bowen}.

Equations~(\ref{eqn:sumamp}) and (\ref{eqn:diffamp}), with the inclusion of squeezing at $\theta_{\rm sq}=0$ ($\theta_{\rm LO}=0$), can be written as
\begin{eqnarray}
  \Tilde{i}_I(\phi=0)&\propto&\frac{1}{2}\left(\Tilde{X}_{1}^{-\Delta}+\Tilde{X}_{1}^{+\Delta}+\Tilde{S}_{1}^{-\Delta}+\Tilde{S}_{1}^{+\Delta}\right),\label{eqn:sqdiffpha}\\
  \Tilde{i}_I(\phi=\pi)&\propto&\frac{1}{2}\left(\Tilde{X}_{1}^{-\Delta}-\Tilde{X}_{1}^{+\Delta}+\Tilde{S}_{1}^{-\Delta}+\Tilde{S}_{1}^{+\Delta}\right).\label{eqn:sqsumpha}
\end{eqnarray}
If the squeezed quadrature angle is misaligned, $\theta_{\rm sq}\neq0$, the measured noise variance will be
\begin{equation}
  V_{\theta_{\rm sq}}=V_{-}\cos^2(\theta_{\rm sq})+V_{+}\sin^2(\theta_{\rm sq}).
\end{equation}
The squeezed quadrature angle changes with the pump field phase. In this work, the pump field phase is manually tuned and not actively controlled.

Experimental loss sets a fundamental limit to the possible improvements in SNR due to squeezing. The expected losses for components of the experiment are summarized in Table~\ref{tab:loss_table}, along with overall experimental efficiencies measured for the upper and lower, signal and squeezing modes. These losses were experimentally measured by propagating bright fields along individual segments of the optical path.
\begin{table}[htb]
  \centering
  \begin{tabular}{c c}
  Component & Efficiency\\
  \midrule
  OPO escape efficiency & $98\pm1\;\%$\\
  Squeezing injection path & $97\pm1\;\%$\\
  Faraday isolator (single pass) & $95\pm1\;\%$\\
  Balanced detector contrast & $97\pm1\;\%$\\
  Photodiode quantum efficiency & $90\pm2\;\%$\\
  Individual interferometer & $95\pm2\;\%$\\
  Combining cavity (single pass) & $96\pm1\;\%$\\
  \midrule
  Signal efficiency, lower (upper) & $77\pm3\;\%$ ($80\pm3\;\%$)\\
  Squeezing mode efficiency, $\eta_1$ ($\eta_2$) & $64\pm3\;\%$ ($69\pm3\;\%$)\\
  Squeezing efficiency, flipper mirror up & $73\pm3\;\%$\\
  \end{tabular}
  \caption{The optical efficiency budget.}\label{tab:loss_table}
\end{table}

\begin{figure}[htb]
  \centering
  \includegraphics[width=\columnwidth]{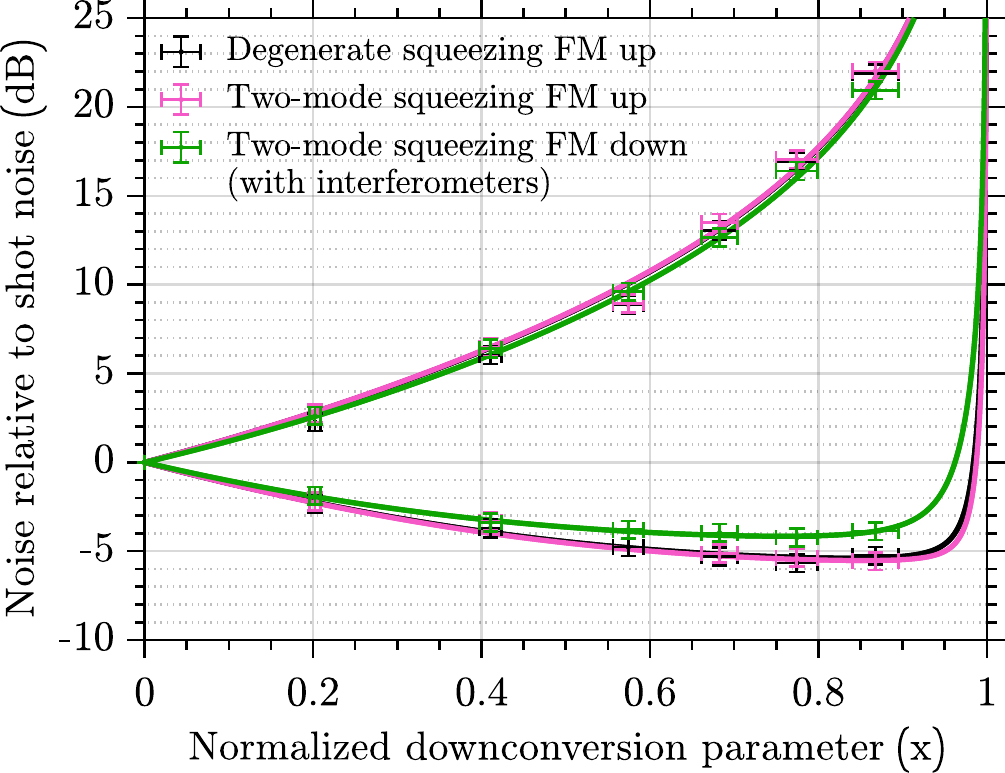}
  \caption{Measurements of squeezing ($\theta_{\rm sq}=0$) and antisqueezing ($\theta_{\rm sq}=\frac{\pi}{2}$) over a range of the normalized downconversion (pump) parameters. Photodetector dark noise has been subtracted from each measurement. We show three configurations, each fitted with Eq.~(\ref{effective dephased measurement}): degenerate squeezing output from the OPO with dc homodyne measurement (black), two-mode squeezing output to balanced detector demodulated at $425\;{\rm MHz}$ (pink), and two-mode squeezing after injection into the twin-interferometers and demodulated at $425\;{\rm MHz}$ (green). Each point represents an average of three measurements over a frequency range of $12\;{\rm kHz}\,\text{--}\,50\;{\rm kHz}$.}\label{fig:SQvNLG}
\end{figure}

First, we characterize the squeezed light source, verifying the expected loss in the process. We measure degenerate and two-mode squeezing with the flipper mirror (FM) up, bypassing the interferometers, as well as characterizing the two-mode squeezed state while operating the experiment as a whole. Both states are produced by the same OPO, which has a threshold of $P_T=66.3\;{\rm mW}$, locked to different pump resonance conditions. Degenerate low-frequency squeezed states are produced when the pump is coresonant with the fundamental resonance, the two-mode high-frequency squeezed state is produced when the pump is operated on its resonance at the midpoint of the OPO fundamental FSR. Measurements of noise power are shown in Fig.~\ref{fig:SQvNLG}, plotted against $x$. Each point represents an average noise power within a shot-noise-limited frequency range $(12\;{\rm kHz}\,\text{--}\,50\;{\rm kHz})$, acquired over three measurements. For the degenerate squeezing measurement, we operate our balanced detector as a dc homodyne detector. 

To model the noise power measurements, the effect of differential loss between the two modes of the squeezed state must be considered. This can be described with a parameter known as dephasing~\cite{Ligo_QResp}. Dephasing acts similarly to phase noise, an effect well understood in degenerate squeezing measurements~\cite{Aoki:06,PhysRevA.67.033802}.

Degradation due to loss and dephasing on a state with squeezed and antisqueezed quadratures, $V_{-}$ and $V_{+}$, results in the degraded noise variance, $V_{\mp}'$~\cite{Ligo_QResp}:
\begin{equation}
  V_{\mp}'=\eta_c\left[(1-\Xi')V_{\mp}+\Xi' V_{\pm}\right]+(1-\eta_c).
  \label{effective dephased measurement}
\end{equation}
Here, $\eta_c$ is the common efficiency,
\begin{equation}
    \eta_c=\frac{\eta_1+\eta_2}{2},
\end{equation}
which depends on $\eta_1$ and $\eta_2$, the respective efficiencies of the lower and upper modes, and $\Xi'$ is the effective dephasing parameter which encompasses the effect of differential loss and phase noise.
\begin{equation}
  \Xi'=\Xi+\theta_{\rm rms}^2-2\Xi \theta_{\rm rms}^2.
\end{equation}
Here, $\theta_{\rm rms}$ is the root-mean-square of the phase fluctuations, and
\begin{equation}
  \Xi=\frac{\eta_1+\eta_2-2\sqrt{\eta_1 \eta_2}}{2(\eta_1+\eta_2)}.
\end{equation}
Figure~\ref{fig:SQvNLG} shows fitted models for each noise power measurement set. The fitting process utilizes Eq.~(\ref{effective dephased measurement}). These models and measurements characterize squeezed state performance and validate our efficiency measurements. These modeling results are summarized in Table~\ref{tab:model_table}.
\begin{table}[htb]
  \ra{1.3}
  \centering
  \begin{tabular}{l c c c}
   Fitted model & Parameter & Value & Error\\
     \midrule
     \multirow{2}{3.2cm}{Degenerate squeezing}&$\eta_c$ & $72\;\%$& $\pm2\;\%$\\
     \cline{2-4}
     &$\Xi'$ & $6\times10^{-5}$&$^{+2.6}_{-0.6}\times10^{-4}$\\
     \midrule
     \multirow{2}{3.2cm}{Two-mode squeezing}&$\eta_c$ & $73\;\%$& $\pm2\;\%$\\
     \cline{2-4}
     &$\Xi'$ & $5\times10^{-5}$&$^{+2.3}_{-0.5}\times10^{-4}$\\
     \midrule
     \multirow{2}{3.2cm}{Two-mode squeezing with interferometers}&$\eta_c$ & $64\;\%$& $\pm1\;\%$\\
     \cline{2-4}
     &$\Xi'$ & $3.7\times10^{-4}$&$^{+1.3}_{-1.3}\times10^{-4}$\\
  \end{tabular}
  \caption{Equation~(\ref{effective dephased measurement}) fitted parameters that produce the models in Fig.~\ref{fig:SQvNLG}.}\label{tab:model_table}
\end{table}

For degenerate and two-mode squeezing measurements, without the dynamics of the interferometers (FM up), the expected effective dephasing will be only due to phase noise, as the loss is uniform at all sideband frequencies, i.e. $\Xi=0$. For the degenerate squeezing measurements, the model returns an effective dephasing corresponding to phase noise of $8\;{\rm mrad}$ with an error range of 0 to $17\;{\rm mrad}$. In the case of two-mode squeezing, it is $6\;{\rm mrad}$ with the bounds of 0 to $14\;{\rm mrad}$. These measurements are consistent with previous characterization of this squeezing source indicating a phase noise of approximately $3\;{\rm mrad}$~\cite{MJYThesis}. The error for the modeled phase noise is relatively large because of the lack of data points close to the oscillation threshold of the OPO. At these higher pump parameters, phase noise limits the SNR improvement~\cite{Aoki:06} and locking of the squeezed quadrature angle is required for accurate averaged measurement. Models for both squeezed states demonstrate a common loss consistent with measurements in Table~\ref{tab:loss_table}.

When characterizing the twin-interferometer system the two-mode squeezed state experienced different losses incurred by the individual interferometers and combining cavity ($\Xi\neq0$). With the returned fitted parameters, and assuming that phase noise is consistent with the prior measurements ($8\;{\rm mrad}$ of phase noise), the remaining dephasing component corresponds with an efficiency of $\eta_1=(62\pm2)\;\%$ and $\eta_2=(66\pm2)\;\%$. These values align with the experimentally measured efficiencies in Table~\ref{tab:loss_table}, considering their uncertainties.

\begin{figure}[htb]
  \centering
  \includegraphics[width=\columnwidth]{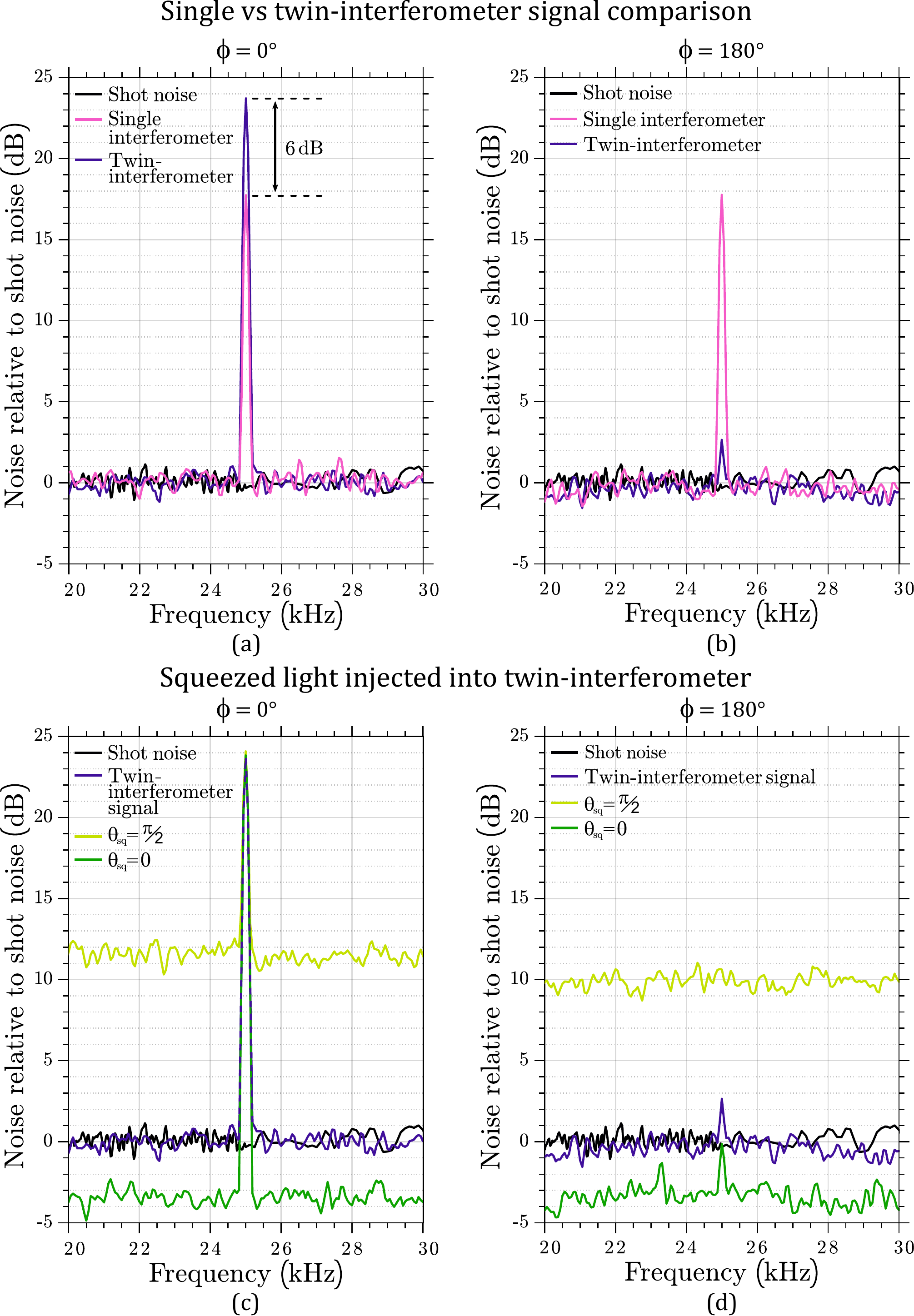}
  \caption{Noise spectra showing the results of the twin-Michelson interferometer experiment normalized to the average shot noise value. The top row of spectra show a $25\;{\rm kHz}$ tone injected into a single interferometer, compared with the same tone simultaneously injected into both interferometers (with $\phi=(0,\pi)$). The bottom row of spectra show the same twin-interferometer tone with the injection of the two-mode squeezed state, operating at a normalized pump parameter of $x=0.65$, corresponding to a squeezing parameter of $r=1.55$.}\label{fig:Amplitude Spec}
\end{figure}

As a second step, an $\Omega=25\;{\rm kHz}$ signal was injected into the twin-interferometer system to characterize the SNR, and demonstrate the evasion of the $3\;{\rm dB}$ heterodyne penalty. Figure~\ref{fig:Amplitude Spec} shows measurements of shot-noise-limited spectra around the injected signal and quantum-enhanced spectra when injecting the two-mode squeezed state. Figure~\ref{fig:Amplitude Spec}(a) and~\ref{fig:Amplitude Spec}(b) compare signal strengths between a single interferometer and the twin interferometer. Figure~\ref{fig:Amplitude Spec}(c) and~\ref{fig:Amplitude Spec}(d) demonstrate the quantum enhancement of the twin-interferometer signals when the two-mode squeezed state is injected. For both nonsqueezing and squeezing injection measurements, we show the sum [$\theta_{\rm LO}=0,\;\phi=0$, Eq.~(\ref{eqn:sumamp})] and difference [$\theta_{\rm LO}=0,\;\phi=\pi$, Eq.~(\ref{eqn:diffamp})] condition of the Michelsons' amplitude quadratures. In our tabletop environment, the lower audio-band frequencies below $10\;{\rm kHz}$ were dominated by acoustic noise and vibrational modes coupled in via mirror mounts, cavities, and other optical elements. The signal and quantum enhancement technique we demonstrate has no fundamental limit in its low-frequency application, other than those technical noise challenges faced across similar experiments~\cite{Stefszky_2012,McKenzie2004,Vahlbruch2007}.

When the amplitude quadratures are coherently added ($\theta_{\rm LO}=0,\;\phi=0$), the signal strength increases by $6\;{\rm dB}$, seen in Fig.~\ref{fig:Amplitude Spec}(a). This is a result of having equal optical power across the two interferometers, such that when the second interferometer is introduced we have both doubled the total carrier power in the system ($+3\;{\rm dB}$) and introduced a dual-sided signal sideband structure ($+3\;{\rm dB}$). The latter explicitly demonstrates the successful experimental circumvention of the conventional $3\;{\rm dB}$ penalty associated with heterodyne readout.

Taking the difference of the amplitude quadratures ($\theta_{\rm LO}=0,\;\phi=\pi$), we coherently cancel the injected signal, a point of interest for correlated interferometry experiments. In Fig.~\ref{fig:Amplitude Spec}(b) we see a coherent signal suppression of more than $15\;{\rm dB}$. Limits to coherent cancellation include carrier power balancing, interferometer lock points, error in signal calibration, the tuning of $\theta_{\rm LO}$ and $\phi$ as well as any drifts in these values over time. In Fig.~\ref{fig:Amplitude Spec}(d) this suppression increases to $17.5\;{\rm dB}$, associated with drifts in the aforementioned parameters and the reduced shot noise level.

When we inject our two-mode squeezed state into the twin-interferometer system, we see either a reduced or increased overall noise floor level, depending on the squeezed quadrature angle. In Fig.~\ref{fig:Amplitude Spec}(c) we directly observe a $3.5\;{\rm dB}$ average reduction when squeezing our readout quadrature ($\theta_{\rm sq}=0$), and an $11.5\;{\rm dB}$ increase in the noise floor when antisqueeze it ($\theta_{\rm sq}=\frac{\pi}{2}$). In Fig.~\ref{fig:Amplitude Spec}(d) we see a similar reduction when $\theta_{\rm sq}=0$ however only a $10\;{\rm dB}$ increase with $\theta_{\rm sq}=\frac{\pi}{2}$. These values, although varying, are consistent with the normalized pump parameter of $x=0.65$, and long timescale pum\textit{p}-power drifts of $5\;\%$ across the measurement period.

The presented results demonstrate a quantum-enhanced, dual Michelson sensor that employs balanced heterodyne readout. We show that a high-frequency quadrature-entangled, two-mode squeezed state can improve the SNR of audio-band signals. Our readout scheme circumvents the conventional $3\;{\rm dB}$ penalty associated with heterodyne readout and can simultaneously obtain quantum-enhanced sum-amplitude and difference-phase quadrature information without introducing additional photodetectors. This flexibility opens opportunities for mixing and filtering amplitude and phase quadratures. Previous experiments~\cite{QD_metrology2013,QD_metrology2016,PhysRevX.7.021008} highlight these opportunities, for technical noise filtering and backaction evasion. Our scheme can provide these opportunities without additional signal loss. This proof-of-principle experiment underlines the potential for high-frequency two-mode squeezing, applicable not only to precision metrology, of which we highlight gravitational wave detection, dark matter detection and quantum gravity tests, but also wavelength-multiplexed quantum communication. It experimentally demonstrates the SNR equivalence of balanced homodyne readout to dual-carrier heterodyne readout.
 
\section*{Data availability}
The data that supports the plots within this Letter and other findings of this study are available from the corresponding author upon reasonable request.

\section*{Acknowledgements}
\begin{acknowledgments}
The authors wish to thank Moritz Mehmet for useful discussions and insights. This research was supported by the Australian Research Council under the Discovery Grant scheme, Grant number DP220102755. D. W. Gould would like to acknowledge the funding from the Australian Government Research Training Program (RTP) Scholarship for his research. V. B. Adya would like to acknowledge the support and funding from the Swedish Research Council (VR starting Grant 2023-0519 and Optical Quantum Sensing environment Grant 2016-06122) and the Wallenberg Center for Quantum Technology (WACQT) in Sweden. The authors would like to also thank the support of the Australian Research Council under the ARC Centre of Excellence for Gravitational Wave Discovery, Grant number CE170100004, and Deutsche Forschungsgemeinschaft (Excellence PhoenixD (EXC 2122, Project ID 390833453), Excellence Quantum Frontiers (EXC 2123, Project ID 390837967)).
\end{acknowledgments}

\section*{Competing Interests}
The authors declare no competing interests.


\begin{thebibliography}{51}%
  \makeatletter
  \providecommand \@ifxundefined [1]{%
   \@ifx{#1\undefined}
  }%
  \providecommand \@ifnum [1]{%
   \ifnum #1\expandafter \@firstoftwo
   \else \expandafter \@secondoftwo
   \fi
  }%
  \providecommand \@ifx [1]{%
   \ifx #1\expandafter \@firstoftwo
   \else \expandafter \@secondoftwo
   \fi
  }%
  \providecommand \natexlab [1]{#1}%
  \providecommand \enquote  [1]{``#1''}%
  \providecommand \bibnamefont  [1]{#1}%
  \providecommand \bibfnamefont [1]{#1}%
  \providecommand \citenamefont [1]{#1}%
  \providecommand \href@noop [0]{\@secondoftwo}%
  \providecommand \href [0]{\begingroup \@sanitize@url \@href}%
  \providecommand \@href[1]{\@@startlink{#1}\@@href}%
  \providecommand \@@href[1]{\endgroup#1\@@endlink}%
  \providecommand \@sanitize@url [0]{\catcode `\\12\catcode `\$12\catcode
    `\&12\catcode `\#12\catcode `\^12\catcode `\_12\catcode `\%12\relax}%
  \providecommand \@@startlink[1]{}%
  \providecommand \@@endlink[0]{}%
  \providecommand \url  [0]{\begingroup\@sanitize@url \@url }%
  \providecommand \@url [1]{\endgroup\@href {#1}{\urlprefix }}%
  \providecommand \urlprefix  [0]{URL }%
  \providecommand \Eprint [0]{\href }%
  \providecommand \doibase [0]{https://doi.org/}%
  \providecommand \selectlanguage [0]{\@gobble}%
  \providecommand \bibinfo  [0]{\@secondoftwo}%
  \providecommand \bibfield  [0]{\@secondoftwo}%
  \providecommand \translation [1]{[#1]}%
  \providecommand \BibitemOpen [0]{}%
  \providecommand \bibitemStop [0]{}%
  \providecommand \bibitemNoStop [0]{.\EOS\space}%
  \providecommand \EOS [0]{\spacefactor3000\relax}%
  \providecommand \BibitemShut  [1]{\csname bibitem#1\endcsname}%
  \let\auto@bib@innerbib\@empty
  \bibitem [{\citenamefont {McCuller}\ \emph {et~al.}(2021)\citenamefont
    {McCuller}, \citenamefont {Dwyer}, \citenamefont {Green}, \citenamefont {Yu},
    \citenamefont {Kuns}, \citenamefont {Barsotti}, \citenamefont {Blair},
    \citenamefont {Brown}, \citenamefont {Effler}, \citenamefont {Evans} \emph
    {et~al.}}]{Ligo_QResp}%
    \BibitemOpen
    \bibfield  {author} {\bibinfo {author} {\bibfnamefont {L.}~\bibnamefont
    {McCuller}}, \bibinfo {author} {\bibfnamefont {S.~E.}\ \bibnamefont {Dwyer}},
    \bibinfo {author} {\bibfnamefont {A.~C.}\ \bibnamefont {Green}}, \bibinfo
    {author} {\bibfnamefont {H.}~\bibnamefont {Yu}}, \bibinfo {author}
    {\bibfnamefont {K.}~\bibnamefont {Kuns}}, \bibinfo {author} {\bibfnamefont
    {L.}~\bibnamefont {Barsotti}}, \bibinfo {author} {\bibfnamefont {C.~D.}\
    \bibnamefont {Blair}}, \bibinfo {author} {\bibfnamefont {D.~D.}\ \bibnamefont
    {Brown}}, \bibinfo {author} {\bibfnamefont {A.}~\bibnamefont {Effler}},
    \bibinfo {author} {\bibfnamefont {M.}~\bibnamefont {Evans}}, \emph {et~al.},\
    }\bibfield  {title} {\bibinfo {title} {{LIGO}'s quantum response to squeezed
    states},\ }\href {https://doi.org/10.1103/PhysRevD.104.062006} {\bibfield
    {journal} {\bibinfo  {journal} {Phys. Rev. D}\ }\textbf {\bibinfo {volume}
    {104}},\ \bibinfo {pages} {062006} (\bibinfo {year} {2021})}\BibitemShut
    {NoStop}%
  \bibitem [{\citenamefont {Ganapathy}\ \emph {et~al.}(2023)\citenamefont
    {Ganapathy}, \citenamefont {Jia}, \citenamefont {Nakano}, \citenamefont {Xu},
    \citenamefont {Aritomi}, \citenamefont {Cullen}, \citenamefont {Kijbunchoo},
    \citenamefont {Dwyer}, \citenamefont {Mullavey}, \citenamefont {McCuller}
    \emph {et~al.}}]{Ganapathy2023}%
    \BibitemOpen
    \bibfield  {author} {\bibinfo {author} {\bibfnamefont {D.}~\bibnamefont
    {Ganapathy}}, \bibinfo {author} {\bibfnamefont {W.}~\bibnamefont {Jia}},
    \bibinfo {author} {\bibfnamefont {M.}~\bibnamefont {Nakano}}, \bibinfo
    {author} {\bibfnamefont {V.}~\bibnamefont {Xu}}, \bibinfo {author}
    {\bibfnamefont {N.}~\bibnamefont {Aritomi}}, \bibinfo {author} {\bibfnamefont
    {T.}~\bibnamefont {Cullen}}, \bibinfo {author} {\bibfnamefont
    {N.}~\bibnamefont {Kijbunchoo}}, \bibinfo {author} {\bibfnamefont {S.~E.}\
    \bibnamefont {Dwyer}}, \bibinfo {author} {\bibfnamefont {A.}~\bibnamefont
    {Mullavey}}, \bibinfo {author} {\bibfnamefont {L.}~\bibnamefont {McCuller}},
    \emph {et~al.} (\bibinfo {collaboration} {LIGO O4 Detector Collaboration}),\
    }\bibfield  {title} {\bibinfo {title} {Broadband quantum enhancement of the
    {LIGO} detectors with frequency-dependent squeezing},\ }\href
    {https://doi.org/10.1103/PhysRevX.13.041021} {\bibfield  {journal} {\bibinfo
    {journal} {Phys. Rev. X}\ }\textbf {\bibinfo {volume} {13}},\ \bibinfo
    {pages} {041021} (\bibinfo {year} {2023})}\BibitemShut {NoStop}%
  \bibitem [{\citenamefont {Acernese}\ \emph {et~al.}(2019)\citenamefont
    {Acernese}, \citenamefont {Agathos}, \citenamefont {Aiello}, \citenamefont
    {Allocca}, \citenamefont {Amato}, \citenamefont {Ansoldi}, \citenamefont
    {Antier}, \citenamefont {Ar\`ene}, \citenamefont {Arnaud}, \citenamefont
    {Ascenzi} \emph {et~al.}}]{PhysRevLett.123.231108}%
    \BibitemOpen
    \bibfield  {author} {\bibinfo {author} {\bibfnamefont {F.}~\bibnamefont
    {Acernese}}, \bibinfo {author} {\bibfnamefont {M.}~\bibnamefont {Agathos}},
    \bibinfo {author} {\bibfnamefont {L.}~\bibnamefont {Aiello}}, \bibinfo
    {author} {\bibfnamefont {A.}~\bibnamefont {Allocca}}, \bibinfo {author}
    {\bibfnamefont {A.}~\bibnamefont {Amato}}, \bibinfo {author} {\bibfnamefont
    {S.}~\bibnamefont {Ansoldi}}, \bibinfo {author} {\bibfnamefont
    {S.}~\bibnamefont {Antier}}, \bibinfo {author} {\bibfnamefont
    {M.}~\bibnamefont {Ar\`ene}}, \bibinfo {author} {\bibfnamefont
    {N.}~\bibnamefont {Arnaud}}, \bibinfo {author} {\bibfnamefont
    {S.}~\bibnamefont {Ascenzi}}, \emph {et~al.} (\bibinfo {collaboration} {Virgo
    Collaboration}),\ }\bibfield  {title} {\bibinfo {title} {Increasing the
    astrophysical reach of the {A}dvanced {V}irgo detector via the application of
    squeezed vacuum states of light},\ }\href
    {https://doi.org/10.1103/PhysRevLett.123.231108} {\bibfield  {journal}
    {\bibinfo  {journal} {Phys. Rev. Lett.}\ }\textbf {\bibinfo {volume} {123}},\
    \bibinfo {pages} {231108} (\bibinfo {year} {2019})}\BibitemShut {NoStop}%
  \bibitem [{\citenamefont {Lough}\ \emph {et~al.}(2021)\citenamefont {Lough},
    \citenamefont {Schreiber}, \citenamefont {Bergamin}, \citenamefont {Grote},
    \citenamefont {Mehmet}, \citenamefont {Vahlbruch}, \citenamefont {Affeldt},
    \citenamefont {Brinkmann}, \citenamefont {Bisht}, \citenamefont {Kringel}
    \emph {et~al.}}]{PhysRevLett.126.041102}%
    \BibitemOpen
    \bibfield  {author} {\bibinfo {author} {\bibfnamefont {J.}~\bibnamefont
    {Lough}}, \bibinfo {author} {\bibfnamefont {E.}~\bibnamefont {Schreiber}},
    \bibinfo {author} {\bibfnamefont {F.}~\bibnamefont {Bergamin}}, \bibinfo
    {author} {\bibfnamefont {H.}~\bibnamefont {Grote}}, \bibinfo {author}
    {\bibfnamefont {M.}~\bibnamefont {Mehmet}}, \bibinfo {author} {\bibfnamefont
    {H.}~\bibnamefont {Vahlbruch}}, \bibinfo {author} {\bibfnamefont
    {C.}~\bibnamefont {Affeldt}}, \bibinfo {author} {\bibfnamefont
    {M.}~\bibnamefont {Brinkmann}}, \bibinfo {author} {\bibfnamefont
    {A.}~\bibnamefont {Bisht}}, \bibinfo {author} {\bibfnamefont
    {V.}~\bibnamefont {Kringel}}, \emph {et~al.},\ }\bibfield  {title} {\bibinfo
    {title} {First demonstration of 6 {dB} quantum noise reduction in a kilometer
    scale gravitational wave observatory},\ }\href
    {https://doi.org/10.1103/PhysRevLett.126.041102} {\bibfield  {journal}
    {\bibinfo  {journal} {Phys. Rev. Lett.}\ }\textbf {\bibinfo {volume} {126}},\
    \bibinfo {pages} {041102} (\bibinfo {year} {2021})}\BibitemShut {NoStop}%
  \bibitem [{\citenamefont {Troullinou}\ \emph {et~al.}(2021)\citenamefont
    {Troullinou}, \citenamefont {Jim\'enez-Mart\'{\i}nez}, \citenamefont {Kong},
    \citenamefont {Lucivero},\ and\ \citenamefont
    {Mitchell}}]{PhysRevLett.127.193601}%
    \BibitemOpen
    \bibfield  {author} {\bibinfo {author} {\bibfnamefont {C.}~\bibnamefont
    {Troullinou}}, \bibinfo {author} {\bibfnamefont {R.}~\bibnamefont
    {Jim\'enez-Mart\'{\i}nez}}, \bibinfo {author} {\bibfnamefont
    {J.}~\bibnamefont {Kong}}, \bibinfo {author} {\bibfnamefont {V.~G.}\
    \bibnamefont {Lucivero}},\ and\ \bibinfo {author} {\bibfnamefont {M.~W.}\
    \bibnamefont {Mitchell}},\ }\bibfield  {title} {\bibinfo {title}
    {Squeezed-light enhancement and backaction evasion in a high sensitivity
    optically pumped magnetometer},\ }\href
    {https://doi.org/10.1103/PhysRevLett.127.193601} {\bibfield  {journal}
    {\bibinfo  {journal} {Phys. Rev. Lett.}\ }\textbf {\bibinfo {volume} {127}},\
    \bibinfo {pages} {193601} (\bibinfo {year} {2021})}\BibitemShut {NoStop}%
  \bibitem [{\citenamefont {Casacio}\ \emph {et~al.}(2021)\citenamefont
    {Casacio}, \citenamefont {Madsen}, \citenamefont {Terrasson}, \citenamefont
    {Waleed}, \citenamefont {Barnscheidt}, \citenamefont {Hage}, \citenamefont
    {Taylor},\ and\ \citenamefont {Bowen}}]{Bowen21}%
    \BibitemOpen
    \bibfield  {author} {\bibinfo {author} {\bibfnamefont {C.~A.}\ \bibnamefont
    {Casacio}}, \bibinfo {author} {\bibfnamefont {L.~S.}\ \bibnamefont {Madsen}},
    \bibinfo {author} {\bibfnamefont {A.}~\bibnamefont {Terrasson}}, \bibinfo
    {author} {\bibfnamefont {M.}~\bibnamefont {Waleed}}, \bibinfo {author}
    {\bibfnamefont {K.}~\bibnamefont {Barnscheidt}}, \bibinfo {author}
    {\bibfnamefont {B.}~\bibnamefont {Hage}}, \bibinfo {author} {\bibfnamefont
    {M.~A.}\ \bibnamefont {Taylor}},\ and\ \bibinfo {author} {\bibfnamefont
    {W.~P.}\ \bibnamefont {Bowen}},\ }\bibfield  {title} {\bibinfo {title}
    {Quantum-enhanced nonlinear microscopy},\ }\href
    {https://doi.org/10.1038/s41586-021-03528-w} {\bibfield  {journal} {\bibinfo
    {journal} {Nature}\ }\textbf {\bibinfo {volume} {594}},\ \bibinfo {pages}
    {201–206} (\bibinfo {year} {2021})}\BibitemShut {NoStop}%
  \bibitem [{\citenamefont {Fan}\ \emph {et~al.}(2015)\citenamefont {Fan},
    \citenamefont {Lawrie},\ and\ \citenamefont {Pooser}}]{PhysRevA.92.053812}%
    \BibitemOpen
    \bibfield  {author} {\bibinfo {author} {\bibfnamefont {W.}~\bibnamefont
    {Fan}}, \bibinfo {author} {\bibfnamefont {B.~J.}\ \bibnamefont {Lawrie}},\
    and\ \bibinfo {author} {\bibfnamefont {R.~C.}\ \bibnamefont {Pooser}},\
    }\bibfield  {title} {\bibinfo {title} {Quantum plasmonic sensing},\ }\href
    {https://doi.org/10.1103/PhysRevA.92.053812} {\bibfield  {journal} {\bibinfo
    {journal} {Phys. Rev. A}\ }\textbf {\bibinfo {volume} {92}},\ \bibinfo
    {pages} {053812} (\bibinfo {year} {2015})}\BibitemShut {NoStop}%
  \bibitem [{\citenamefont {Armstrong}\ \emph {et~al.}(2012)\citenamefont
    {Armstrong}, \citenamefont {Morizur}, \citenamefont {Janousek}, \citenamefont
    {Hage}, \citenamefont {Treps}, \citenamefont {Lam},\ and\ \citenamefont
    {Bachor}}]{armstrong2012programmable}%
    \BibitemOpen
    \bibfield  {author} {\bibinfo {author} {\bibfnamefont {S.}~\bibnamefont
    {Armstrong}}, \bibinfo {author} {\bibfnamefont {J.-F.}\ \bibnamefont
    {Morizur}}, \bibinfo {author} {\bibfnamefont {J.}~\bibnamefont {Janousek}},
    \bibinfo {author} {\bibfnamefont {B.}~\bibnamefont {Hage}}, \bibinfo {author}
    {\bibfnamefont {N.}~\bibnamefont {Treps}}, \bibinfo {author} {\bibfnamefont
    {P.~K.}\ \bibnamefont {Lam}},\ and\ \bibinfo {author} {\bibfnamefont {H.-A.}\
    \bibnamefont {Bachor}},\ }\bibfield  {title} {\bibinfo {title} {Programmable
    multimode quantum networks},\ }\href {https://doi.org/10.1038/ncomms2033}
    {\bibfield  {journal} {\bibinfo  {journal} {Nature communications}\ }\textbf
    {\bibinfo {volume} {3}},\ \bibinfo {pages} {1026} (\bibinfo {year}
    {2012})}\BibitemShut {NoStop}%
  \bibitem [{\citenamefont {Collett}\ \emph {et~al.}(1987)\citenamefont
    {Collett}, \citenamefont {Loudon},\ and\ \citenamefont {Gardiner}}]{Collett}%
    \BibitemOpen
    \bibfield  {author} {\bibinfo {author} {\bibfnamefont {M.}~\bibnamefont
    {Collett}}, \bibinfo {author} {\bibfnamefont {R.}~\bibnamefont {Loudon}},\
    and\ \bibinfo {author} {\bibfnamefont {C.~W.}\ \bibnamefont {Gardiner}},\
    }\bibfield  {title} {\bibinfo {title} {Quantum theory of optical homodyne and
    heterodyne detection},\ }\href {https://doi.org/10.1080/09500348714550811}
    {\bibfield  {journal} {\bibinfo  {journal} {Journal of Modern Optics}\
    }\textbf {\bibinfo {volume} {34}},\ \bibinfo {pages} {881} (\bibinfo {year}
    {1987})}\BibitemShut {NoStop}%
  \bibitem [{\citenamefont {Yuen}\ and\ \citenamefont {Chan}(1983)}]{Yuen:83}%
    \BibitemOpen
    \bibfield  {author} {\bibinfo {author} {\bibfnamefont {H.~P.}\ \bibnamefont
    {Yuen}}\ and\ \bibinfo {author} {\bibfnamefont {V.~W.~S.}\ \bibnamefont
    {Chan}},\ }\bibfield  {title} {\bibinfo {title} {Noise in homodyne and
    heterodyne detection},\ }\href {https://doi.org/10.1364/OL.8.000177}
    {\bibfield  {journal} {\bibinfo  {journal} {Opt. Lett.}\ }\textbf {\bibinfo
    {volume} {8}},\ \bibinfo {pages} {177} (\bibinfo {year} {1983})}\BibitemShut
    {NoStop}%
  \bibitem [{\citenamefont {Meylahn}\ \emph {et~al.}(2022)\citenamefont
    {Meylahn}, \citenamefont {Willke},\ and\ \citenamefont
    {Vahlbruch}}]{PhysRevLett.129.121103}%
    \BibitemOpen
    \bibfield  {author} {\bibinfo {author} {\bibfnamefont {F.}~\bibnamefont
    {Meylahn}}, \bibinfo {author} {\bibfnamefont {B.}~\bibnamefont {Willke}},\
    and\ \bibinfo {author} {\bibfnamefont {H.}~\bibnamefont {Vahlbruch}},\
    }\bibfield  {title} {\bibinfo {title} {Squeezed states of light for future
    gravitational wave detectors at a wavelength of 1550 nm},\ }\href
    {https://doi.org/10.1103/PhysRevLett.129.121103} {\bibfield  {journal}
    {\bibinfo  {journal} {Phys. Rev. Lett.}\ }\textbf {\bibinfo {volume} {129}},\
    \bibinfo {pages} {121103} (\bibinfo {year} {2022})}\BibitemShut {NoStop}%
  \bibitem [{\citenamefont {Fritschel}\ \emph {et~al.}(2014)\citenamefont
    {Fritschel}, \citenamefont {Evans},\ and\ \citenamefont
    {Frolov}}]{Fritschel:14}%
    \BibitemOpen
    \bibfield  {author} {\bibinfo {author} {\bibfnamefont {P.}~\bibnamefont
    {Fritschel}}, \bibinfo {author} {\bibfnamefont {M.}~\bibnamefont {Evans}},\
    and\ \bibinfo {author} {\bibfnamefont {V.}~\bibnamefont {Frolov}},\
    }\bibfield  {title} {\bibinfo {title} {Balanced homodyne readout for quantum
    limited gravitational wave detectors},\ }\href
    {https://doi.org/10.1364/OE.22.004224} {\bibfield  {journal} {\bibinfo
    {journal} {Opt. Express}\ }\textbf {\bibinfo {volume} {22}},\ \bibinfo
    {pages} {4224} (\bibinfo {year} {2014})}\BibitemShut {NoStop}%
  \bibitem [{\citenamefont {Stefszky}\ \emph {et~al.}(2012)\citenamefont
    {Stefszky}, \citenamefont {Mow-Lowry}, \citenamefont {Chua}, \citenamefont
    {Shaddock}, \citenamefont {Buchler}, \citenamefont {Vahlbruch}, \citenamefont
    {Khalaidovski}, \citenamefont {Schnabel}, \citenamefont {Lam},\ and\
    \citenamefont {McClelland}}]{Stefszky_2012}%
    \BibitemOpen
    \bibfield  {author} {\bibinfo {author} {\bibfnamefont {M.~S.}\ \bibnamefont
    {Stefszky}}, \bibinfo {author} {\bibfnamefont {C.~M.}\ \bibnamefont
    {Mow-Lowry}}, \bibinfo {author} {\bibfnamefont {S.~S.~Y.}\ \bibnamefont
    {Chua}}, \bibinfo {author} {\bibfnamefont {D.~A.}\ \bibnamefont {Shaddock}},
    \bibinfo {author} {\bibfnamefont {B.~C.}\ \bibnamefont {Buchler}}, \bibinfo
    {author} {\bibfnamefont {H.}~\bibnamefont {Vahlbruch}}, \bibinfo {author}
    {\bibfnamefont {A.}~\bibnamefont {Khalaidovski}}, \bibinfo {author}
    {\bibfnamefont {R.}~\bibnamefont {Schnabel}}, \bibinfo {author}
    {\bibfnamefont {P.~K.}\ \bibnamefont {Lam}},\ and\ \bibinfo {author}
    {\bibfnamefont {D.~E.}\ \bibnamefont {McClelland}},\ }\bibfield  {title}
    {\bibinfo {title} {Balanced homodyne detection of optical quantum states at
    audio-band frequencies and below},\ }\href
    {https://doi.org/10.1088/0264-9381/29/14/145015} {\bibfield  {journal}
    {\bibinfo  {journal} {Classical and Quantum Gravity}\ }\textbf {\bibinfo
    {volume} {29}},\ \bibinfo {pages} {145015} (\bibinfo {year}
    {2012})}\BibitemShut {NoStop}%
  \bibitem [{\citenamefont {Steinlechner}\ \emph {et~al.}(2015)\citenamefont
    {Steinlechner}, \citenamefont {Barr}, \citenamefont {Bell}, \citenamefont
    {Danilishin}, \citenamefont {Gl\"afke}, \citenamefont {Gr\"af}, \citenamefont
    {Hennig}, \citenamefont {Houston}, \citenamefont {Huttner}, \citenamefont
    {Leavey} \emph {et~al.}}]{Steinlechner2015}%
    \BibitemOpen
    \bibfield  {author} {\bibinfo {author} {\bibfnamefont {S.}~\bibnamefont
    {Steinlechner}}, \bibinfo {author} {\bibfnamefont {B.~W.}\ \bibnamefont
    {Barr}}, \bibinfo {author} {\bibfnamefont {A.~S.}\ \bibnamefont {Bell}},
    \bibinfo {author} {\bibfnamefont {S.~L.}\ \bibnamefont {Danilishin}},
    \bibinfo {author} {\bibfnamefont {A.}~\bibnamefont {Gl\"afke}}, \bibinfo
    {author} {\bibfnamefont {C.}~\bibnamefont {Gr\"af}}, \bibinfo {author}
    {\bibfnamefont {J.-S.}\ \bibnamefont {Hennig}}, \bibinfo {author}
    {\bibfnamefont {E.~A.}\ \bibnamefont {Houston}}, \bibinfo {author}
    {\bibfnamefont {S.~H.}\ \bibnamefont {Huttner}}, \bibinfo {author}
    {\bibfnamefont {S.~S.}\ \bibnamefont {Leavey}}, \emph {et~al.},\ }\bibfield
    {title} {\bibinfo {title} {Local-oscillator noise coupling in balanced
    homodyne readout for advanced gravitational wave detectors},\ }\href
    {https://doi.org/10.1103/PhysRevD.92.072009} {\bibfield  {journal} {\bibinfo
    {journal} {Phys. Rev. D}\ }\textbf {\bibinfo {volume} {92}},\ \bibinfo
    {pages} {072009} (\bibinfo {year} {2015})}\BibitemShut {NoStop}%
  \bibitem [{\citenamefont {McKenzie}\ \emph {et~al.}(2004)\citenamefont
    {McKenzie}, \citenamefont {Grosse}, \citenamefont {Bowen}, \citenamefont
    {Whitcomb}, \citenamefont {Gray}, \citenamefont {McClelland},\ and\
    \citenamefont {Lam}}]{McKenzie2004}%
    \BibitemOpen
    \bibfield  {author} {\bibinfo {author} {\bibfnamefont {K.}~\bibnamefont
    {McKenzie}}, \bibinfo {author} {\bibfnamefont {N.}~\bibnamefont {Grosse}},
    \bibinfo {author} {\bibfnamefont {W.~P.}\ \bibnamefont {Bowen}}, \bibinfo
    {author} {\bibfnamefont {S.~E.}\ \bibnamefont {Whitcomb}}, \bibinfo {author}
    {\bibfnamefont {M.~B.}\ \bibnamefont {Gray}}, \bibinfo {author}
    {\bibfnamefont {D.~E.}\ \bibnamefont {McClelland}},\ and\ \bibinfo {author}
    {\bibfnamefont {P.~K.}\ \bibnamefont {Lam}},\ }\bibfield  {title} {\bibinfo
    {title} {Squeezing in the audio gravitational-wave detection band},\ }\href
    {https://doi.org/10.1103/PhysRevLett.93.161105} {\bibfield  {journal}
    {\bibinfo  {journal} {Phys. Rev. Lett.}\ }\textbf {\bibinfo {volume} {93}},\
    \bibinfo {pages} {161105} (\bibinfo {year} {2004})}\BibitemShut {NoStop}%
  \bibitem [{\citenamefont {Vahlbruch}\ \emph {et~al.}(2007)\citenamefont
    {Vahlbruch}, \citenamefont {Chelkowski}, \citenamefont {Danzmann},\ and\
    \citenamefont {Schnabel}}]{Vahlbruch2007}%
    \BibitemOpen
    \bibfield  {author} {\bibinfo {author} {\bibfnamefont {H.}~\bibnamefont
    {Vahlbruch}}, \bibinfo {author} {\bibfnamefont {S.}~\bibnamefont
    {Chelkowski}}, \bibinfo {author} {\bibfnamefont {K.}~\bibnamefont
    {Danzmann}},\ and\ \bibinfo {author} {\bibfnamefont {R.}~\bibnamefont
    {Schnabel}},\ }\bibfield  {title} {\bibinfo {title} {Quantum engineering of
    squeezed states for quantum communication and metrology},\ }\href
    {https://doi.org/10.1088/1367-2630/9/10/371} {\bibfield  {journal} {\bibinfo
    {journal} {New Journal of Physics}\ }\textbf {\bibinfo {volume} {9}},\
    \bibinfo {pages} {371} (\bibinfo {year} {2007})}\BibitemShut {NoStop}%
  \bibitem [{\citenamefont {Vahlbruch}\ \emph {et~al.}(2006)\citenamefont
    {Vahlbruch}, \citenamefont {Chelkowski}, \citenamefont {Hage}, \citenamefont
    {Franzen}, \citenamefont {Danzmann},\ and\ \citenamefont
    {Schnabel}}]{PhysRevLett.97.011101}%
    \BibitemOpen
    \bibfield  {author} {\bibinfo {author} {\bibfnamefont {H.}~\bibnamefont
    {Vahlbruch}}, \bibinfo {author} {\bibfnamefont {S.}~\bibnamefont
    {Chelkowski}}, \bibinfo {author} {\bibfnamefont {B.}~\bibnamefont {Hage}},
    \bibinfo {author} {\bibfnamefont {A.}~\bibnamefont {Franzen}}, \bibinfo
    {author} {\bibfnamefont {K.}~\bibnamefont {Danzmann}},\ and\ \bibinfo
    {author} {\bibfnamefont {R.}~\bibnamefont {Schnabel}},\ }\bibfield  {title}
    {\bibinfo {title} {Coherent control of vacuum squeezing in the
    gravitational-wave detection band},\ }\href
    {https://doi.org/10.1103/PhysRevLett.97.011101} {\bibfield  {journal}
    {\bibinfo  {journal} {Phys. Rev. Lett.}\ }\textbf {\bibinfo {volume} {97}},\
    \bibinfo {pages} {011101} (\bibinfo {year} {2006})}\BibitemShut {NoStop}%
  \bibitem [{\citenamefont {Spollard}\ \emph {et~al.}(2021)\citenamefont
    {Spollard}, \citenamefont {Roberts}, \citenamefont {Sambridge}, \citenamefont
    {McKenzie},\ and\ \citenamefont {Shaddock}}]{Spollard:21}%
    \BibitemOpen
    \bibfield  {author} {\bibinfo {author} {\bibfnamefont {J.~T.}\ \bibnamefont
    {Spollard}}, \bibinfo {author} {\bibfnamefont {L.~E.}\ \bibnamefont
    {Roberts}}, \bibinfo {author} {\bibfnamefont {C.~S.}\ \bibnamefont
    {Sambridge}}, \bibinfo {author} {\bibfnamefont {K.}~\bibnamefont
    {McKenzie}},\ and\ \bibinfo {author} {\bibfnamefont {D.~A.}\ \bibnamefont
    {Shaddock}},\ }\bibfield  {title} {\bibinfo {title} {Mitigation of phase
    noise and doppler-induced frequency offsets in coherent random amplitude
    modulated continuou\textit{s}-wave lidar},\ }\href {https://doi.org/10.1364/OE.416128}
    {\bibfield  {journal} {\bibinfo  {journal} {Opt. Express}\ }\textbf {\bibinfo
    {volume} {29}},\ \bibinfo {pages} {9060} (\bibinfo {year}
    {2021})}\BibitemShut {NoStop}%
  \bibitem [{\citenamefont {Yu}\ \emph {et~al.}(2023)\citenamefont {Yu},
    \citenamefont {Gewecke}, \citenamefont {S\"{u}dbeck}, \citenamefont
    {Sch\"{o}nbeck}, \citenamefont {Schnabel},\ and\ \citenamefont
    {Rembe}}]{Yu:23}%
    \BibitemOpen
    \bibfield  {author} {\bibinfo {author} {\bibfnamefont {M.}~\bibnamefont
    {Yu}}, \bibinfo {author} {\bibfnamefont {P.}~\bibnamefont {Gewecke}},
    \bibinfo {author} {\bibfnamefont {J.}~\bibnamefont {S\"{u}dbeck}}, \bibinfo
    {author} {\bibfnamefont {A.}~\bibnamefont {Sch\"{o}nbeck}}, \bibinfo {author}
    {\bibfnamefont {R.}~\bibnamefont {Schnabel}},\ and\ \bibinfo {author}
    {\bibfnamefont {C.}~\bibnamefont {Rembe}},\ }\bibfield  {title} {\bibinfo
    {title} {Heterodyne laser doppler vibrometer with squeezed light
    enhancement},\ }\href {https://doi.org/10.1364/OL.494064} {\bibfield
    {journal} {\bibinfo  {journal} {Opt. Lett.}\ }\textbf {\bibinfo {volume}
    {48}},\ \bibinfo {pages} {5607} (\bibinfo {year} {2023})}\BibitemShut
    {NoStop}%
  \bibitem [{\citenamefont {de~Andrade}\ \emph {et~al.}(2020)\citenamefont
    {de~Andrade}, \citenamefont {Kerdoncuff}, \citenamefont {Berg-S{\o}rensen},
    \citenamefont {Gehring}, \citenamefont {Lassen},\ and\ \citenamefont
    {Andersen}}]{deAndrade:20}%
    \BibitemOpen
    \bibfield  {author} {\bibinfo {author} {\bibfnamefont {R.~B.}\ \bibnamefont
    {de~Andrade}}, \bibinfo {author} {\bibfnamefont {H.}~\bibnamefont
    {Kerdoncuff}}, \bibinfo {author} {\bibfnamefont {K.}~\bibnamefont
    {Berg-S{\o}rensen}}, \bibinfo {author} {\bibfnamefont {T.}~\bibnamefont
    {Gehring}}, \bibinfo {author} {\bibfnamefont {M.}~\bibnamefont {Lassen}},\
    and\ \bibinfo {author} {\bibfnamefont {U.~L.}\ \bibnamefont {Andersen}},\
    }\bibfield  {title} {\bibinfo {title} {Quantum-enhanced continuou\textit{s}-wave
    stimulated raman scattering spectroscopy},\ }\href
    {https://doi.org/10.1364/OPTICA.386584} {\bibfield  {journal} {\bibinfo
    {journal} {Optica}\ }\textbf {\bibinfo {volume} {7}},\ \bibinfo {pages} {470}
    (\bibinfo {year} {2020})}\BibitemShut {NoStop}%
  \bibitem [{\citenamefont {Drever}\ \emph {et~al.}(1983)\citenamefont {Drever},
    \citenamefont {Hall}, \citenamefont {Kowalski}, \citenamefont {Hough},
    \citenamefont {Ford}, \citenamefont {Munley},\ and\ \citenamefont
    {Ward}}]{PDH}%
    \BibitemOpen
    \bibfield  {author} {\bibinfo {author} {\bibfnamefont {R.~W.~P.}\
    \bibnamefont {Drever}}, \bibinfo {author} {\bibfnamefont {J.~L.}\
    \bibnamefont {Hall}}, \bibinfo {author} {\bibfnamefont {F.~V.}\ \bibnamefont
    {Kowalski}}, \bibinfo {author} {\bibfnamefont {J.}~\bibnamefont {Hough}},
    \bibinfo {author} {\bibfnamefont {G.~M.}\ \bibnamefont {Ford}}, \bibinfo
    {author} {\bibfnamefont {A.~J.}\ \bibnamefont {Munley}},\ and\ \bibinfo
    {author} {\bibfnamefont {H.}~\bibnamefont {Ward}},\ }\bibfield  {title}
    {\bibinfo {title} {Laser phase and frequency stabilization using an optical
    resonator},\ }\href {https://doi.org/10.1007/BF00702605} {\bibfield
    {journal} {\bibinfo  {journal} {Applied Physics B}\ }\textbf {\bibinfo
    {volume} {31}},\ \bibinfo {pages} {97} (\bibinfo {year} {1983})}\BibitemShut
    {NoStop}%
  \bibitem [{\citenamefont {B\'eguin}\ \emph {et~al.}(2014)\citenamefont
    {B\'eguin}, \citenamefont {Bookjans}, \citenamefont {Christensen},
    \citenamefont {S\o{}rensen}, \citenamefont {M\"uller}, \citenamefont
    {Polzik},\ and\ \citenamefont {Appel}}]{PRL_1D_OL}%
    \BibitemOpen
    \bibfield  {author} {\bibinfo {author} {\bibfnamefont {J.-B.}\ \bibnamefont
    {B\'eguin}}, \bibinfo {author} {\bibfnamefont {E.~M.}\ \bibnamefont
    {Bookjans}}, \bibinfo {author} {\bibfnamefont {S.~L.}\ \bibnamefont
    {Christensen}}, \bibinfo {author} {\bibfnamefont {H.~L.}\ \bibnamefont
    {S\o{}rensen}}, \bibinfo {author} {\bibfnamefont {J.~H.}\ \bibnamefont
    {M\"uller}}, \bibinfo {author} {\bibfnamefont {E.~S.}\ \bibnamefont
    {Polzik}},\ and\ \bibinfo {author} {\bibfnamefont {J.}~\bibnamefont
    {Appel}},\ }\bibfield  {title} {\bibinfo {title} {Generation and detection of
    a sub-poissonian atom number distribution in a one-dimensional optical
    lattice},\ }\href {https://doi.org/10.1103/PhysRevLett.113.263603} {\bibfield
     {journal} {\bibinfo  {journal} {Phys. Rev. Lett.}\ }\textbf {\bibinfo
    {volume} {113}},\ \bibinfo {pages} {263603} (\bibinfo {year}
    {2014})}\BibitemShut {NoStop}%
  \bibitem [{\citenamefont {Xie}\ \emph {et~al.}(2018)\citenamefont {Xie},
    \citenamefont {Yang},\ and\ \citenamefont {Feng}}]{Xie:18}%
    \BibitemOpen
    \bibfield  {author} {\bibinfo {author} {\bibfnamefont {B.}~\bibnamefont
    {Xie}}, \bibinfo {author} {\bibfnamefont {P.}~\bibnamefont {Yang}},\ and\
    \bibinfo {author} {\bibfnamefont {S.}~\bibnamefont {Feng}},\ }\bibfield
    {title} {\bibinfo {title} {Phase-sensitive heterodyne detection of two-mode
    squeezed light without noise penalty},\ }\href
    {https://doi.org/10.1364/JOSAB.35.002342} {\bibfield  {journal} {\bibinfo
    {journal} {J. Opt. Soc. Am. B}\ }\textbf {\bibinfo {volume} {35}},\ \bibinfo
    {pages} {2342} (\bibinfo {year} {2018})}\BibitemShut {NoStop}%
  \bibitem [{\citenamefont {Zhang}\ \emph {et~al.}(2021)\citenamefont {Zhang},
    \citenamefont {Jones}, \citenamefont {Smetana}, \citenamefont {Miao},
    \citenamefont {Martynov}, \citenamefont {Freise},\ and\ \citenamefont
    {Ballmer}}]{Zhang_two_carrier}%
    \BibitemOpen
    \bibfield  {author} {\bibinfo {author} {\bibfnamefont {T.}~\bibnamefont
    {Zhang}}, \bibinfo {author} {\bibfnamefont {P.}~\bibnamefont {Jones}},
    \bibinfo {author} {\bibfnamefont {J.}~\bibnamefont {Smetana}}, \bibinfo
    {author} {\bibfnamefont {H.}~\bibnamefont {Miao}}, \bibinfo {author}
    {\bibfnamefont {D.}~\bibnamefont {Martynov}}, \bibinfo {author}
    {\bibfnamefont {A.}~\bibnamefont {Freise}},\ and\ \bibinfo {author}
    {\bibfnamefont {S.~W.}\ \bibnamefont {Ballmer}},\ }\bibfield  {title}
    {\bibinfo {title} {Two-carrier scheme: Evading the 3 {dB} quantum penalty of
    heterodyne readout in gravitational-wave detectors},\ }\href
    {https://doi.org/10.1103/PhysRevLett.126.221301} {\bibfield  {journal}
    {\bibinfo  {journal} {Phys. Rev. Lett.}\ }\textbf {\bibinfo {volume} {126}},\
    \bibinfo {pages} {221301} (\bibinfo {year} {2021})}\BibitemShut {NoStop}%
  \bibitem [{\citenamefont {Zhang}(2021)}]{JZ_two_carrier}%
    \BibitemOpen
    \bibfield  {author} {\bibinfo {author} {\bibfnamefont {J.}~\bibnamefont
    {Zhang}},\ }\emph {\bibinfo {title} {Opto-mechanical interactions in laser
    interferometric gravitational wave detectors}},\ \href
    {https://doi.org/10.26182/bejn-d793} {Ph.D. thesis},\ \bibinfo  {school} {The
    University of Western Australia} (\bibinfo {year} {2021})\BibitemShut
    {NoStop}%
  \bibitem [{\citenamefont {Vermeulen}\ \emph {et~al.}(2021)\citenamefont
    {Vermeulen}, \citenamefont {Aiello}, \citenamefont {Ejlli}, \citenamefont
    {Griffiths}, \citenamefont {James}, \citenamefont {Dooley},\ and\
    \citenamefont {Grote}}]{Vermeulen_2021}%
    \BibitemOpen
    \bibfield  {author} {\bibinfo {author} {\bibfnamefont {S.~M.}\ \bibnamefont
    {Vermeulen}}, \bibinfo {author} {\bibfnamefont {L.}~\bibnamefont {Aiello}},
    \bibinfo {author} {\bibfnamefont {A.}~\bibnamefont {Ejlli}}, \bibinfo
    {author} {\bibfnamefont {W.~L.}\ \bibnamefont {Griffiths}}, \bibinfo {author}
    {\bibfnamefont {A.~L.}\ \bibnamefont {James}}, \bibinfo {author}
    {\bibfnamefont {K.~L.}\ \bibnamefont {Dooley}},\ and\ \bibinfo {author}
    {\bibfnamefont {H.}~\bibnamefont {Grote}},\ }\bibfield  {title} {\bibinfo
    {title} {An experiment for observing quantum gravity phenomena using twin
    table-top {3D} interferometers},\ }\href
    {https://doi.org/10.1088/1361-6382/abe757} {\bibfield  {journal} {\bibinfo
    {journal} {Classical and Quantum Gravity}\ }\textbf {\bibinfo {volume}
    {38}},\ \bibinfo {pages} {085008} (\bibinfo {year} {2021})}\BibitemShut
    {NoStop}%
  \bibitem [{\citenamefont {Chou}\ \emph {et~al.}(2017)\citenamefont {Chou},
    \citenamefont {Glass}, \citenamefont {Gustafson}, \citenamefont {Hogan},
    \citenamefont {Kamai}, \citenamefont {Kwon}, \citenamefont {Lanza},
    \citenamefont {McCuller}, \citenamefont {Meyer}, \citenamefont {Richardson}
    \emph {et~al.}}]{TheHolometer}%
    \BibitemOpen
    \bibfield  {author} {\bibinfo {author} {\bibfnamefont {A.}~\bibnamefont
    {Chou}}, \bibinfo {author} {\bibfnamefont {H.}~\bibnamefont {Glass}},
    \bibinfo {author} {\bibfnamefont {H.~R.}\ \bibnamefont {Gustafson}}, \bibinfo
    {author} {\bibfnamefont {C.}~\bibnamefont {Hogan}}, \bibinfo {author}
    {\bibfnamefont {B.~L.}\ \bibnamefont {Kamai}}, \bibinfo {author}
    {\bibfnamefont {O.}~\bibnamefont {Kwon}}, \bibinfo {author} {\bibfnamefont
    {R.}~\bibnamefont {Lanza}}, \bibinfo {author} {\bibfnamefont
    {L.}~\bibnamefont {McCuller}}, \bibinfo {author} {\bibfnamefont {S.~S.}\
    \bibnamefont {Meyer}}, \bibinfo {author} {\bibfnamefont {J.}~\bibnamefont
    {Richardson}}, \emph {et~al.},\ }\bibfield  {title} {\bibinfo {title} {The
    holometer: an instrument to probe planckian quantum geometry},\ }\href
    {https://doi.org/10.1088/1361-6382/aa5e5c} {\bibfield  {journal} {\bibinfo
    {journal} {Classical and Quantum Gravity}\ }\textbf {\bibinfo {volume}
    {34}},\ \bibinfo {pages} {065005} (\bibinfo {year} {2017})}\BibitemShut
    {NoStop}%
  \bibitem [{\citenamefont {Pradyumna}\ \emph {et~al.}(2020)\citenamefont
    {Pradyumna}, \citenamefont {Losero}, \citenamefont {Ruo-Berchera},
    \citenamefont {Traina}, \citenamefont {Zucco}, \citenamefont {Jacobsen},
    \citenamefont {Andersen}, \citenamefont {Degiovanni}, \citenamefont
    {Genovese},\ and\ \citenamefont {Gehring}}]{Pradyumna2020}%
    \BibitemOpen
    \bibfield  {author} {\bibinfo {author} {\bibfnamefont {S.~T.}\ \bibnamefont
    {Pradyumna}}, \bibinfo {author} {\bibfnamefont {E.}~\bibnamefont {Losero}},
    \bibinfo {author} {\bibfnamefont {I.}~\bibnamefont {Ruo-Berchera}}, \bibinfo
    {author} {\bibfnamefont {P.}~\bibnamefont {Traina}}, \bibinfo {author}
    {\bibfnamefont {M.}~\bibnamefont {Zucco}}, \bibinfo {author} {\bibfnamefont
    {C.~S.}\ \bibnamefont {Jacobsen}}, \bibinfo {author} {\bibfnamefont {U.~L.}\
    \bibnamefont {Andersen}}, \bibinfo {author} {\bibfnamefont {I.~P.}\
    \bibnamefont {Degiovanni}}, \bibinfo {author} {\bibfnamefont
    {M.}~\bibnamefont {Genovese}},\ and\ \bibinfo {author} {\bibfnamefont
    {T.}~\bibnamefont {Gehring}},\ }\bibfield  {title} {\bibinfo {title} {Twin
    beam quantum-enhanced correlated interferometry for testing fundamental
    physics},\ }\href {https://doi.org/10.1038/s42005-020-0368-5} {\bibfield
    {journal} {\bibinfo  {journal} {Communications Physics}\ }\textbf {\bibinfo
    {volume} {3}},\ \bibinfo {pages} {104} (\bibinfo {year} {2020})}\BibitemShut
    {NoStop}%
\bibitem [{\citenamefont {Ejlli}\ \emph {et~al.}(2023)\citenamefont {Ejlli},
    \citenamefont {Vermeulen}, \citenamefont {Schwartz}, \citenamefont {Aiello},\
    and\ \citenamefont {Grote}}]{DMPolarimetryEjlli}%
    \BibitemOpen
    \bibfield  {author} {\bibinfo {author} {\bibfnamefont {A.}~\bibnamefont
    {Ejlli}}, \bibinfo {author} {\bibfnamefont {S.~M.}\ \bibnamefont
    {Vermeulen}}, \bibinfo {author} {\bibfnamefont {E.}~\bibnamefont {Schwartz}},
    \bibinfo {author} {\bibfnamefont {L.}~\bibnamefont {Aiello}},\ and\ \bibinfo
    {author} {\bibfnamefont {H.}~\bibnamefont {Grote}},\ }\bibfield  {title}
    {\bibinfo {title} {Probing dark matter with polarimetry techniques},\ }\href
    {https://doi.org/10.1103/PhysRevD.107.083035} {\bibfield  {journal} {\bibinfo
     {journal} {Physical Review D: Particles and Fields}\ }\textbf {\bibinfo
    {volume} {107}},\ \bibinfo {pages} {083035} (\bibinfo {year}
    {2023})}\BibitemShut {NoStop}%
  \bibitem [{\citenamefont {Heinze}\ \emph {et~al.}(2024)\citenamefont {Heinze},
    \citenamefont {Gill}, \citenamefont {Dmitriev}, \citenamefont {Smetana}, \citenamefont {Yan}, \citenamefont {Boyer},
    \citenamefont {Martynov},\ and\ \citenamefont {Evans}}]{LIDAFirstResults}%
    \BibitemOpen
    \bibfield  {author} {\bibinfo {author} {\bibfnamefont {J.}~\bibnamefont
    {Heinze}}, \bibinfo {author} {\bibfnamefont {A.}~\bibnamefont {Gill}},
    \bibinfo {author} {\bibfnamefont {A.}~\bibnamefont {Dmitriev}}, \bibinfo
    {author} {\bibfnamefont {J.}~\bibnamefont {Smetana}}, \bibinfo
    {author} {\bibfnamefont {T.}~\bibnamefont {Yan}}, \bibinfo {author}
    {\bibfnamefont {V.}~\bibnamefont {Boyer}}, \bibinfo {author} {\bibfnamefont
    {D.}~\bibnamefont {Martynov}},\ and\ \bibinfo {author} {\bibfnamefont
    {M.}~\bibnamefont {Evans}},\ }\bibfield  {title} {\bibinfo {title} {First
    results of the laser-interferometric detector for axions ({{LIDA}})},\ }\href
    {https://doi.org/10.1103/PhysRevLett.132.191002} {\bibfield  {journal}
    {\bibinfo  {journal} {Physical Review Letters}\ }\textbf {\bibinfo {volume}
    {132}},\ \bibinfo {pages} {191002} (\bibinfo {year} {2024})}\BibitemShut
    {NoStop}%
  \bibitem [{\citenamefont {Ruo~Berchera}\ \emph {et~al.}(2013)\citenamefont
    {Ruo~Berchera}, \citenamefont {Degiovanni}, \citenamefont {Olivares},\ and\
    \citenamefont {Genovese}}]{QL_coupled_int}%
    \BibitemOpen
    \bibfield  {author} {\bibinfo {author} {\bibfnamefont {I.}~\bibnamefont
    {Ruo~Berchera}}, \bibinfo {author} {\bibfnamefont {I.~P.}\ \bibnamefont
    {Degiovanni}}, \bibinfo {author} {\bibfnamefont {S.}~\bibnamefont
    {Olivares}},\ and\ \bibinfo {author} {\bibfnamefont {M.}~\bibnamefont
    {Genovese}},\ }\bibfield  {title} {\bibinfo {title} {Quantum light in coupled
    interferometers for quantum gravity tests},\ }\href
    {https://doi.org/10.1103/PhysRevLett.110.213601} {\bibfield  {journal}
    {\bibinfo  {journal} {Phys. Rev. Lett.}\ }\textbf {\bibinfo {volume} {110}},\
    \bibinfo {pages} {213601} (\bibinfo {year} {2013})}\BibitemShut {NoStop}%
  \bibitem [{\citenamefont {Ruo-Berchera}\ \emph {et~al.}(2015)\citenamefont
    {Ruo-Berchera}, \citenamefont {Degiovanni}, \citenamefont {Olivares},
    \citenamefont {Samantaray}, \citenamefont {Traina},\ and\ \citenamefont
    {Genovese}}]{one_two_SQ_corr_int}%
    \BibitemOpen
    \bibfield  {author} {\bibinfo {author} {\bibfnamefont {I.}~\bibnamefont
    {Ruo-Berchera}}, \bibinfo {author} {\bibfnamefont {I.~P.}\ \bibnamefont
    {Degiovanni}}, \bibinfo {author} {\bibfnamefont {S.}~\bibnamefont
    {Olivares}}, \bibinfo {author} {\bibfnamefont {N.}~\bibnamefont
    {Samantaray}}, \bibinfo {author} {\bibfnamefont {P.}~\bibnamefont {Traina}},\
    and\ \bibinfo {author} {\bibfnamefont {M.}~\bibnamefont {Genovese}},\
    }\bibfield  {title} {\bibinfo {title} {One- and two-mode squeezed light in
    correlated interferometry},\ }\href
    {https://doi.org/10.1103/PhysRevA.92.053821} {\bibfield  {journal} {\bibinfo
    {journal} {Phys. Rev. A}\ }\textbf {\bibinfo {volume} {92}},\ \bibinfo
    {pages} {053821} (\bibinfo {year} {2015})}\BibitemShut {NoStop}%
  \bibitem [{\citenamefont {S{\"u}dbeck}\ \emph {et~al.}(2020)\citenamefont
    {S{\"u}dbeck}, \citenamefont {Steinlechner}, \citenamefont {Korobko},\ and\
    \citenamefont {Schnabel}}]{EPR_Sudbeck}%
    \BibitemOpen
    \bibfield  {author} {\bibinfo {author} {\bibfnamefont {J.}~\bibnamefont
    {S{\"u}dbeck}}, \bibinfo {author} {\bibfnamefont {S.}~\bibnamefont
    {Steinlechner}}, \bibinfo {author} {\bibfnamefont {M.}~\bibnamefont
    {Korobko}},\ and\ \bibinfo {author} {\bibfnamefont {R.}~\bibnamefont
    {Schnabel}},\ }\bibfield  {title} {\bibinfo {title} {Demonstration of
    interferometer enhancement through {E}instein--{P}odolsky--{R}osen
    entanglement},\ }\href {https://doi.org/10.1038/s41566-019-0583-3} {\bibfield
     {journal} {\bibinfo  {journal} {Nature Photonics}\ }\textbf {\bibinfo
    {volume} {14}},\ \bibinfo {pages} {240} (\bibinfo {year} {2020})}\BibitemShut
    {NoStop}%
  \bibitem [{\citenamefont {Yap}\ \emph {et~al.}(2020)\citenamefont {Yap},
    \citenamefont {Altin}, \citenamefont {McRae}, \citenamefont {Slagmolen},
    \citenamefont {Ward},\ and\ \citenamefont {McClelland}}]{EPR_Yap}%
    \BibitemOpen
    \bibfield  {author} {\bibinfo {author} {\bibfnamefont {M.~J.}\ \bibnamefont
    {Yap}}, \bibinfo {author} {\bibfnamefont {P.}~\bibnamefont {Altin}}, \bibinfo
    {author} {\bibfnamefont {T.~G.}\ \bibnamefont {McRae}}, \bibinfo {author}
    {\bibfnamefont {B.~J.~J.}\ \bibnamefont {Slagmolen}}, \bibinfo {author}
    {\bibfnamefont {R.~L.}\ \bibnamefont {Ward}},\ and\ \bibinfo {author}
    {\bibfnamefont {D.~E.}\ \bibnamefont {McClelland}},\ }\bibfield  {title}
    {\bibinfo {title} {Generation and control of frequency-dependent squeezing
    via einstein--podolsky--rosen entanglement},\ }\href
    {https://doi.org/10.1038/s41566-019-0582-4} {\bibfield  {journal} {\bibinfo
    {journal} {Nature Photonics}\ }\textbf {\bibinfo {volume} {14}},\ \bibinfo
    {pages} {223} (\bibinfo {year} {2020})}\BibitemShut {NoStop}%
  \bibitem [{\citenamefont {Gould}\ \emph {et~al.}(2021)\citenamefont {Gould},
    \citenamefont {Yap}, \citenamefont {Adya}, \citenamefont {Slagmolen},
    \citenamefont {Ward},\ and\ \citenamefont {McClelland}}]{Gould_QNC}%
    \BibitemOpen
    \bibfield  {author} {\bibinfo {author} {\bibfnamefont {D.~W.}\ \bibnamefont
    {Gould}}, \bibinfo {author} {\bibfnamefont {M.~J.}\ \bibnamefont {Yap}},
    \bibinfo {author} {\bibfnamefont {V.~B.}\ \bibnamefont {Adya}}, \bibinfo
    {author} {\bibfnamefont {B.~J.~J.}\ \bibnamefont {Slagmolen}}, \bibinfo
    {author} {\bibfnamefont {R.~L.}\ \bibnamefont {Ward}},\ and\ \bibinfo
    {author} {\bibfnamefont {D.~E.}\ \bibnamefont {McClelland}},\ }\bibfield
    {title} {\bibinfo {title} {Optimal quantum noise cancellation with an
    entangled witness channel},\ }\href
    {https://doi.org/10.1103/PhysRevResearch.3.043079} {\bibfield  {journal}
    {\bibinfo  {journal} {Phys. Rev. Res.}\ }\textbf {\bibinfo {volume} {3}},\
    \bibinfo {pages} {043079} (\bibinfo {year} {2021})}\BibitemShut {NoStop}%
  \bibitem [{\citenamefont {Zhuang}\ \emph {et~al.}(2018)\citenamefont {Zhuang},
    \citenamefont {Zhang},\ and\ \citenamefont {Shapiro}}]{PhysRevA.97.032329}%
    \BibitemOpen
    \bibfield  {author} {\bibinfo {author} {\bibfnamefont {Q.}~\bibnamefont
    {Zhuang}}, \bibinfo {author} {\bibfnamefont {Z.}~\bibnamefont {Zhang}},\ and\
    \bibinfo {author} {\bibfnamefont {J.~H.}\ \bibnamefont {Shapiro}},\
    }\bibfield  {title} {\bibinfo {title} {Distributed quantum sensing using
    continuou\textit{s}-variable multipartite entanglement},\ }\href
    {https://doi.org/10.1103/PhysRevA.97.032329} {\bibfield  {journal} {\bibinfo
    {journal} {Phys. Rev. A}\ }\textbf {\bibinfo {volume} {97}},\ \bibinfo
    {pages} {032329} (\bibinfo {year} {2018})}\BibitemShut {NoStop}%
  \bibitem [{\citenamefont {Song}\ \emph {et~al.}(2014)\citenamefont {Song},
    \citenamefont {Yonezawa}, \citenamefont {Kuntz}, \citenamefont {Heurs},\ and\
    \citenamefont {Huntington}}]{Qtele}%
    \BibitemOpen
    \bibfield  {author} {\bibinfo {author} {\bibfnamefont {H.}~\bibnamefont
    {Song}}, \bibinfo {author} {\bibfnamefont {H.}~\bibnamefont {Yonezawa}},
    \bibinfo {author} {\bibfnamefont {K.~B.}\ \bibnamefont {Kuntz}}, \bibinfo
    {author} {\bibfnamefont {M.}~\bibnamefont {Heurs}},\ and\ \bibinfo {author}
    {\bibfnamefont {E.~H.}\ \bibnamefont {Huntington}},\ }\bibfield  {title}
    {\bibinfo {title} {Quantum teleportation in space and frequency using
    entangled pairs of photons from a frequency comb},\ }\href
    {https://doi.org/10.1103/PhysRevA.90.042337} {\bibfield  {journal} {\bibinfo
    {journal} {Phys. Rev. A}\ }\textbf {\bibinfo {volume} {90}},\ \bibinfo
    {pages} {042337} (\bibinfo {year} {2014})}\BibitemShut {NoStop}%
  \bibitem [{\citenamefont {Chou}\ \emph {et~al.}(2016)\citenamefont {Chou},
    \citenamefont {Gustafson}, \citenamefont {Hogan}, \citenamefont {Kamai},
    \citenamefont {Kwon}, \citenamefont {Lanza}, \citenamefont {McCuller},
    \citenamefont {Meyer}, \citenamefont {Richardson}, \citenamefont {Stoughton}
    \emph {et~al.}}]{Holometer1stmeas}%
    \BibitemOpen
    \bibfield  {author} {\bibinfo {author} {\bibfnamefont {A.~S.}\ \bibnamefont
    {Chou}}, \bibinfo {author} {\bibfnamefont {R.}~\bibnamefont {Gustafson}},
    \bibinfo {author} {\bibfnamefont {C.}~\bibnamefont {Hogan}}, \bibinfo
    {author} {\bibfnamefont {B.}~\bibnamefont {Kamai}}, \bibinfo {author}
    {\bibfnamefont {O.}~\bibnamefont {Kwon}}, \bibinfo {author} {\bibfnamefont
    {R.}~\bibnamefont {Lanza}}, \bibinfo {author} {\bibfnamefont
    {L.}~\bibnamefont {McCuller}}, \bibinfo {author} {\bibfnamefont {S.~S.}\
    \bibnamefont {Meyer}}, \bibinfo {author} {\bibfnamefont {J.}~\bibnamefont
    {Richardson}}, \bibinfo {author} {\bibfnamefont {C.}~\bibnamefont
    {Stoughton}}, \emph {et~al.} (\bibinfo {collaboration} {Holometer
    Collaboration}),\ }\bibfield  {title} {\bibinfo {title} {First measurements
    of high frequency cros\textit{s}-spectra from a pair of large michelson
    interferometers},\ }\href {https://doi.org/10.1103/PhysRevLett.117.111102}
    {\bibfield  {journal} {\bibinfo  {journal} {Phys. Rev. Lett.}\ }\textbf
    {\bibinfo {volume} {117}},\ \bibinfo {pages} {111102} (\bibinfo {year}
    {2016})}\BibitemShut {NoStop}%
  \bibitem [{\citenamefont {Lee}\ \emph {et~al.}(2011)\citenamefont {Lee},
    \citenamefont {Benichi}, \citenamefont {Takeno}, \citenamefont {Takeda},
    \citenamefont {Webb}, \citenamefont {Huntington},\ and\ \citenamefont
    {Furusawa}}]{Qtele_Science}%
    \BibitemOpen
    \bibfield  {author} {\bibinfo {author} {\bibfnamefont {N.}~\bibnamefont
    {Lee}}, \bibinfo {author} {\bibfnamefont {H.}~\bibnamefont {Benichi}},
    \bibinfo {author} {\bibfnamefont {Y.}~\bibnamefont {Takeno}}, \bibinfo
    {author} {\bibfnamefont {S.}~\bibnamefont {Takeda}}, \bibinfo {author}
    {\bibfnamefont {J.}~\bibnamefont {Webb}}, \bibinfo {author} {\bibfnamefont
    {E.}~\bibnamefont {Huntington}},\ and\ \bibinfo {author} {\bibfnamefont
    {A.}~\bibnamefont {Furusawa}},\ }\bibfield  {title} {\bibinfo {title}
    {Teleportation of nonclassical wave packets of light},\ }\href
    {https://doi.org/10.1126/science.1201034} {\bibfield  {journal} {\bibinfo
    {journal} {Science}\ }\textbf {\bibinfo {volume} {332}},\ \bibinfo {pages}
    {330} (\bibinfo {year} {2011})}\BibitemShut
    {NoStop}%
  \bibitem [{\citenamefont {Chapman}\ \emph {et~al.}(2023)\citenamefont
    {Chapman}, \citenamefont {Miloshevsky}, \citenamefont {Lu}, \citenamefont
    {Rao}, \citenamefont {Alshowkan},\ and\ \citenamefont {Peters}}]{Chapman:23}%
    \BibitemOpen
    \bibfield  {author} {\bibinfo {author} {\bibfnamefont {J.~C.}\ \bibnamefont
    {Chapman}}, \bibinfo {author} {\bibfnamefont {A.}~\bibnamefont
    {Miloshevsky}}, \bibinfo {author} {\bibfnamefont {H.-H.}\ \bibnamefont {Lu}},
    \bibinfo {author} {\bibfnamefont {N.}~\bibnamefont {Rao}}, \bibinfo {author}
    {\bibfnamefont {M.}~\bibnamefont {Alshowkan}},\ and\ \bibinfo {author}
    {\bibfnamefont {N.~A.}\ \bibnamefont {Peters}},\ }\bibfield  {title}
    {\bibinfo {title} {Two-mode squeezing over deployed fiber coexisting with
    conventional communications},\ }\href {https://doi.org/10.1364/OE.492539}
    {\bibfield  {journal} {\bibinfo  {journal} {Opt. Express}\ }\textbf {\bibinfo
    {volume} {31}},\ \bibinfo {pages} {26254} (\bibinfo {year}
    {2023})}\BibitemShut {NoStop}%
  \bibitem [{\citenamefont {Junker}\ \emph {et~al.}(2022)\citenamefont {Junker},
    \citenamefont {Wilken}, \citenamefont {Johny}, \citenamefont {Steinmeyer},\
    and\ \citenamefont {Heurs}}]{Junker2022}%
    \BibitemOpen
    \bibfield  {author} {\bibinfo {author} {\bibfnamefont {J.}~\bibnamefont
    {Junker}}, \bibinfo {author} {\bibfnamefont {D.}~\bibnamefont {Wilken}},
    \bibinfo {author} {\bibfnamefont {N.}~\bibnamefont {Johny}}, \bibinfo
    {author} {\bibfnamefont {D.}~\bibnamefont {Steinmeyer}},\ and\ \bibinfo
    {author} {\bibfnamefont {M.}~\bibnamefont {Heurs}},\ }\bibfield  {title}
    {\bibinfo {title} {Frequency-dependent squeezing from a detuned squeezer},\
    }\href {https://doi.org/10.1103/PhysRevLett.129.033602} {\bibfield  {journal}
    {\bibinfo  {journal} {Phys. Rev. Lett.}\ }\textbf {\bibinfo {volume} {129}},\
    \bibinfo {pages} {033602} (\bibinfo {year} {2022})}\BibitemShut {NoStop}%
  \bibitem [{\citenamefont {Stefszky}\ \emph {et~al.}(2011)\citenamefont
    {Stefszky}, \citenamefont {Mow-Lowry}, \citenamefont {McKenzie},
    \citenamefont {Chua}, \citenamefont {Buchler}, \citenamefont {Symul},
    \citenamefont {McClelland},\ and\ \citenamefont {Lam}}]{StefszkyJOPB2011}%
    \BibitemOpen
    \bibfield  {author} {\bibinfo {author} {\bibfnamefont {M.}~\bibnamefont
    {Stefszky}}, \bibinfo {author} {\bibfnamefont {C.~M.}\ \bibnamefont
    {Mow-Lowry}}, \bibinfo {author} {\bibfnamefont {K.}~\bibnamefont {McKenzie}},
    \bibinfo {author} {\bibfnamefont {S.}~\bibnamefont {Chua}}, \bibinfo {author}
    {\bibfnamefont {B.~C.}\ \bibnamefont {Buchler}}, \bibinfo {author}
    {\bibfnamefont {T.}~\bibnamefont {Symul}}, \bibinfo {author} {\bibfnamefont
    {D.~E.}\ \bibnamefont {McClelland}},\ and\ \bibinfo {author} {\bibfnamefont
    {P.~K.}\ \bibnamefont {Lam}},\ }\bibfield  {title} {\bibinfo {title} {An
    investigation of doubly-resonant optical parametric oscillators and nonlinear
    crystals for squeezing},\ }\href
    {https://doi.org/10.1088/0953-4075/44/1/015502} {\bibfield  {journal}
    {\bibinfo  {journal} {J. Phys. B: At. Mol. Opt. Phys.}\ }\textbf {\bibinfo
    {volume} {44}},\ \bibinfo {pages} {015502} (\bibinfo {year}
    {2011})}\BibitemShut {NoStop}%
  \bibitem [{\citenamefont {Reid}\ and\ \citenamefont
    {Drummond}(1988)}]{QNOPO_Reid}%
    \BibitemOpen
    \bibfield  {author} {\bibinfo {author} {\bibfnamefont {M.~D.}\ \bibnamefont
    {Reid}}\ and\ \bibinfo {author} {\bibfnamefont {P.~D.}\ \bibnamefont
    {Drummond}},\ }\bibfield  {title} {\bibinfo {title} {Quantum correlations of
    phase in nondegenerate parametric oscillation},\ }\href
    {https://doi.org/10.1103/PhysRevLett.60.2731} {\bibfield  {journal} {\bibinfo
     {journal} {Phys. Rev. Lett.}\ }\textbf {\bibinfo {volume} {60}},\ \bibinfo
    {pages} {2731} (\bibinfo {year} {1988})}\BibitemShut {NoStop}%
  \bibitem [{\citenamefont {Bowen}\ \emph {et~al.}(2003)\citenamefont {Bowen},
    \citenamefont {Schnabel}, \citenamefont {Lam},\ and\ \citenamefont
    {Ralph}}]{ExpCriteriaCV_Bowen}%
    \BibitemOpen
    \bibfield  {author} {\bibinfo {author} {\bibfnamefont {W.~P.}\ \bibnamefont
    {Bowen}}, \bibinfo {author} {\bibfnamefont {R.}~\bibnamefont {Schnabel}},
    \bibinfo {author} {\bibfnamefont {P.~K.}\ \bibnamefont {Lam}},\ and\ \bibinfo
    {author} {\bibfnamefont {T.~C.}\ \bibnamefont {Ralph}},\ }\bibfield  {title}
    {\bibinfo {title} {Experimental investigation of criteria for continuous
    variable entanglement},\ }\href
    {https://doi.org/10.1103/PhysRevLett.90.043601} {\bibfield  {journal}
    {\bibinfo  {journal} {Phys. Rev. Lett.}\ }\textbf {\bibinfo {volume} {90}},\
    \bibinfo {pages} {043601} (\bibinfo {year} {2003})}\BibitemShut {NoStop}%
  \bibitem [{\citenamefont {Aoki}\ \emph {et~al.}(2006)\citenamefont {Aoki},
    \citenamefont {Takahashi},\ and\ \citenamefont {Furusawa}}]{Aoki:06}%
    \BibitemOpen
    \bibfield  {author} {\bibinfo {author} {\bibfnamefont {T.}~\bibnamefont
    {Aoki}}, \bibinfo {author} {\bibfnamefont {G.}~\bibnamefont {Takahashi}},\
    and\ \bibinfo {author} {\bibfnamefont {A.}~\bibnamefont {Furusawa}},\
    }\bibfield  {title} {\bibinfo {title} {Squeezing at 946nm with periodically
    poled {KTiOPO$_{4}$}},\ }\href {https://doi.org/10.1364/OE.14.006930}
    {\bibfield  {journal} {\bibinfo  {journal} {Opt. Express}\ }\textbf {\bibinfo
    {volume} {14}},\ \bibinfo {pages} {6930} (\bibinfo {year}
    {2006})}\BibitemShut {NoStop}%
  \bibitem [{\citenamefont {Zhang}\ \emph {et~al.}(2003)\citenamefont {Zhang},
    \citenamefont {Goh}, \citenamefont {Chou}, \citenamefont {Lodahl},\ and\
    \citenamefont {Kimble}}]{PhysRevA.67.033802}%
    \BibitemOpen
    \bibfield  {author} {\bibinfo {author} {\bibfnamefont {T.~C.}\ \bibnamefont
    {Zhang}}, \bibinfo {author} {\bibfnamefont {K.~W.}\ \bibnamefont {Goh}},
    \bibinfo {author} {\bibfnamefont {C.~W.}\ \bibnamefont {Chou}}, \bibinfo
    {author} {\bibfnamefont {P.}~\bibnamefont {Lodahl}},\ and\ \bibinfo {author}
    {\bibfnamefont {H.~J.}\ \bibnamefont {Kimble}},\ }\bibfield  {title}
    {\bibinfo {title} {Quantum teleportation of light beams},\ }\href
    {https://doi.org/10.1103/PhysRevA.67.033802} {\bibfield  {journal} {\bibinfo
    {journal} {Phys. Rev. A}\ }\textbf {\bibinfo {volume} {67}},\ \bibinfo
    {pages} {033802} (\bibinfo {year} {2003})}\BibitemShut {NoStop}%
  \bibitem [{\citenamefont {Yap}(2020)}]{MJYThesis}%
    \BibitemOpen
    \bibfield  {author} {\bibinfo {author} {\bibfnamefont {M.~J.}\ \bibnamefont
    {Yap}},\ }\emph {\bibinfo {title} {Quantum noise reduction for
    gravitational-wave interferometers with non-classical states}},\ \href
    {https://doi.org/10.25911/5f0edb7e55763} {Ph.D. thesis},\ \bibinfo  {school}
    {Australian National University} (\bibinfo {year} {2020})\BibitemShut
    {NoStop}%
  \bibitem [{\citenamefont {Steinlechner}\ \emph {et~al.}(2013)\citenamefont
    {Steinlechner}, \citenamefont {Bauchrowitz}, \citenamefont {Meinders},
    \citenamefont {M{\"u}ller-Ebhardt}, \citenamefont {Danzmann},\ and\
    \citenamefont {Schnabel}}]{QD_metrology2013}%
    \BibitemOpen
    \bibfield  {author} {\bibinfo {author} {\bibfnamefont {S.}~\bibnamefont
    {Steinlechner}}, \bibinfo {author} {\bibfnamefont {J.}~\bibnamefont
    {Bauchrowitz}}, \bibinfo {author} {\bibfnamefont {M.}~\bibnamefont
    {Meinders}}, \bibinfo {author} {\bibfnamefont {H.}~\bibnamefont
    {M{\"u}ller-Ebhardt}}, \bibinfo {author} {\bibfnamefont {K.}~\bibnamefont
    {Danzmann}},\ and\ \bibinfo {author} {\bibfnamefont {R.}~\bibnamefont
    {Schnabel}},\ }\bibfield  {title} {\bibinfo {title} {Quantum-dense
    metrology},\ }\href {https://doi.org/10.1038/nphoton.2013.150} {\bibfield
    {journal} {\bibinfo  {journal} {Nature Photonics}\ }\textbf {\bibinfo
    {volume} {7}},\ \bibinfo {pages} {626} (\bibinfo {year} {2013})}\BibitemShut
    {NoStop}%
  \bibitem [{\citenamefont {Ast}\ \emph {et~al.}(2016)\citenamefont {Ast},
    \citenamefont {Steinlechner},\ and\ \citenamefont
    {Schnabel}}]{QD_metrology2016}%
    \BibitemOpen
    \bibfield  {author} {\bibinfo {author} {\bibfnamefont {M.}~\bibnamefont
    {Ast}}, \bibinfo {author} {\bibfnamefont {S.}~\bibnamefont {Steinlechner}},\
    and\ \bibinfo {author} {\bibfnamefont {R.}~\bibnamefont {Schnabel}},\
    }\bibfield  {title} {\bibinfo {title} {Reduction of classical measurement
    noise via quantum-dense metrology},\ }\href
    {https://doi.org/10.1103/PhysRevLett.117.180801} {\bibfield  {journal}
    {\bibinfo  {journal} {Phys. Rev. Lett.}\ }\textbf {\bibinfo {volume} {117}},\
    \bibinfo {pages} {180801} (\bibinfo {year} {2016})}\BibitemShut {NoStop}%
  \bibitem [{\citenamefont {Kampel}\ \emph {et~al.}(2017)\citenamefont {Kampel},
    \citenamefont {Peterson}, \citenamefont {Fischer}, \citenamefont {Yu},
    \citenamefont {Cicak}, \citenamefont {Simmonds}, \citenamefont {Lehnert},\
    and\ \citenamefont {Regal}}]{PhysRevX.7.021008}%
    \BibitemOpen
    \bibfield  {author} {\bibinfo {author} {\bibfnamefont {N.~S.}\ \bibnamefont
    {Kampel}}, \bibinfo {author} {\bibfnamefont {R.~W.}\ \bibnamefont
    {Peterson}}, \bibinfo {author} {\bibfnamefont {R.}~\bibnamefont {Fischer}},
    \bibinfo {author} {\bibfnamefont {P.-L.}\ \bibnamefont {Yu}}, \bibinfo
    {author} {\bibfnamefont {K.}~\bibnamefont {Cicak}}, \bibinfo {author}
    {\bibfnamefont {R.~W.}\ \bibnamefont {Simmonds}}, \bibinfo {author}
    {\bibfnamefont {K.~W.}\ \bibnamefont {Lehnert}},\ and\ \bibinfo {author}
    {\bibfnamefont {C.~A.}\ \bibnamefont {Regal}},\ }\bibfield  {title} {\bibinfo
    {title} {Improving broadband displacement detection with quantum
    correlations},\ }\href {https://doi.org/10.1103/PhysRevX.7.021008} {\bibfield
     {journal} {\bibinfo  {journal} {Phys. Rev. X}\ }\textbf {\bibinfo {volume}
    {7}},\ \bibinfo {pages} {021008} (\bibinfo {year} {2017})}\BibitemShut
    {NoStop}%
  \end{thebibliography}
\end{document}